\documentclass[11pt,a4paper]{article}
\pdfoutput=1 

\usepackage{jheppub} 

\usepackage[T1]{fontenc} 
\usepackage{slashed}
\usepackage[normalem]{ulem}
\usepackage{xcolor}
\usepackage{tikz}
\usepackage{mathtools}
\usetikzlibrary{decorations.pathmorphing}
\usetikzlibrary{decorations.markings}
\tikzset{snake it/.style={decorate, decoration=snake}}
\tikzset{fermion/.style={decoration={markings,
  mark=at position .5 with {\arrow{>}}},postaction={decorate}}}

\newcommand{\ap}{\alpha'}

\title{Effective interactions of the open bosonic string via field theory}

\author[a]{Lucia M.\ Garozzo,}
\author[b,c]{Alfredo Guevara}

\affiliation[a]{Department of Physics and Astronomy, Uppsala University, \\ Box 516, 75120 Uppsala, Sweden}
\affiliation[b]{Center for the Fundamental Laws of Nature, Harvard University, Cambridge, MA 02138}
\affiliation[c]{Society of Fellows, Harvard University, Cambridge, MA 02138
}

\emailAdd{lucia.garozzo@physics.uu.se, aguevaragonzalez@fas.harvard.edu}

\abstract{We describe a method to extract an effective Lagrangian description for open bosonic strings, at zero transcendentality. The method relies on a particular formulation of its scattering amplitudes derived from color-kinematics duality. More precisely, starting from a $(DF)^2 + \text{YM}$ quantum field theory, we integrate out all the massive degrees of freedom to generate an expansion in the inverse string tension $\ap$. We explicitly compute the Lagrangian terms through $\mathcal{O}(\ap^4)$, and target the sector of operators proportional to $F^4$ to all orders in $\ap$.
}

\preprint{UUITP-08/24 
}  

\begin{document} 
\maketitle
\flushbottom

\section{Introduction}
What properties of string theory are unique to its worldsheet description? Formulating string theory as an effective field theory is essential for answering this question, since it enables understanding it in terms of QFT principles. At the very least one would want to obtain an effective action from which string theory observables, namely scattering amplitudes, can be computed in this language. However, a full low-energy description of string theory in terms of an effective action has remained elusive so far. For the case of abelian and slowly varying fields, the effective action for the bosonic open string \cite{Fradkin:1985qd} and open superstring \cite{Metsaev:1987qp} was found to be reproduced at all orders in the inverse string tension $\ap$ by the Born-Infeld action \cite{Born:1934gh}. In the non-abelian case, a similar all-orders formulation is not known. Ref.~\cite{Tseytlin:1997csa} proposed a non-abelian generalization of the Born-Infeld action, which reproduces a sector of the operators contributing to the open superstring effective action (see also the review~\cite{Tseytlin:1999dj}). This Born-Infeld-like action generates operators proportional to even powers of the field strength $F^{\mu \nu}$ and independent of derivatives, which are only a sector of all the possible operators. Lacking a full all-orders description, efforts so far were based on determining the operators order by order in $\ap$. In regards to the open superstring, the leading term is known to be the Yang-Mills term $F^2$, while terms at order $\mathcal{O}(\ap)$ are absent. The bosonic piece of the $\mathcal{O}(\ap^2)$ action was obtained in ref.~\cite{Tseytlin:1986ti, Gross:1986iv} and the fermionic piece in ref.~\cite{Cederwall:2001td, Bergshoeff:2001dc}. At $\mathcal{O}(\ap^3)$ both bosonic \cite{Koerber:2001uu} and fermionic \cite{Collinucci:2002ac} terms are known as well, while at $\mathcal{O}(\ap^4)$ only the bosonic piece has been computed \cite{Koerber:2002zb, Barreiro:2012aw}. See also refs.~\cite{Chandia:2003sh, Barreiro:2005hv} for progress in computing four and five-point effective actions at all orders in $\ap$.

The progress described above was mainly achieved by writing down all the operators which can contribute at a certain order, and then constraining their coefficients by comparing with known open string scattering amplitudes. In recent years, a new guiding principle for constructing the effective action has emerged under the names of \textit{color-kinematics duality} and \textit{double copy} \cite{Bern:2008qj}. Under this, open string amplitudes satisfy KLT relations that link them directly to field theory amplitudes. Schematically, one can write (see section \ref{sec:review} for precise notation):

\begin{align}
A_{\text{open}}  =& Z \underset{\mathrm{KLT}}{\otimes}
\, A_{\text{SYM}} \\
A_{\text{bos}} =& Z \underset{\mathrm{KLT}}{\otimes}\, B_{\text{YM+$\alpha'$}} \label{eq:bos_ampli}
\end{align}
Here ${\otimes}_{\textrm{KLT}}$ stands for the (field-theory) KLT product defined below. Furthermore, $Z$ stands for worldsheet integrals carrying uniform transcendentality: in their $\alpha'$-expansion, each power matches the transcendentality weight of the multiple zeta values in its coefficient. Finally, $A_{\text{SYM}} $ and $B_{\text{YM+$\alpha'$}}$ are field theory amplitudes: the former corresponds to the familiar Super-Yang-Mills (SYM) theory, whereas the latter extends YM by an infinite tower of higher-derivative operators in the amplitude \cite{Huang:2016tag,Azevedo:2018dgo}. See refs.~\cite{Carrasco:2019yyn, Bonnefoy:2021qgu, Menezes:2021dyp, Chi:2021mio, Carrasco:2021ptp, Carrasco:2022lbm, Carrasco:2023wib, Chen:2023ekh} for some recent developments in the study of higher-derivative operators in the context of color-kinematics duality.

Our ultimate task is to write down a field theory Lagrangian leading to the (tree-level) amplitude $A_{\text{bos}}$ and understand its properties.  The highest transcendentality sector of the bosonic string effective action is obtained by discarding the fermionic pieces of the open superstring effective action. This, however, is not sufficient to reproduce the full bosonic amplitudes as it is immediately manifest at $\mathcal{O}(\ap)$, where the bosonic string amplitude contains terms which arise from an $F^3$ operator and are not proportional to Riemann zeta values. These are precisely the new operators that enter in $B_{\text{YM+$\alpha'$}}$. More precisely, the amplitude $B_{\text{YM+$\alpha'$}}$ contains all the effective operators in the so-called zero-transcendentality sector.

In this paper we aim to recast the zero-transcendentality sector as a quantum field theory in its own right, by writing down its effective description. For this we will follow recent insights from color-kinematics duality. More precisely, it has been pointed out that a particular $(DF)^2 + \rm{YM}$ Lagrangian  \cite{Johansson:2017srf}, involving pure YM with higher-derivative terms coupled to a colored scalar, generates the polarization dependence of open bosonic string amplitudes \cite{Huang:2016tag,Azevedo:2018dgo}. This Lagrangian, however, involves unphysical states outside the bosonic string spectrum. This is explicit in the $(DF)^2$ kinetic term, leading to ghost-like and tachyonic (massive) spin-1 modes, denoted by $B^{\mu}$. The colored scalar, denoted by $\varphi^\alpha$, is also tachyonic, with $m=-1/\ap$. In the amplitude \eqref{eq:bos_ampli}, these poles conspire into the scalar tachyonic poles of the bosonic string. At the level of the Lagrangian, however, in order to obtain an effective description of the scattering of massless external vector states, the massive degrees of freedom need to be integrated out. To achieve this, we recast the Lagrangian such that the degrees of freedom of a massless gauge field are manifestly separated from the rest. The resulting Lagrangian takes the schematic form

\begin{equation}
    \mathcal{L} = -\frac{1}{2 \ap} B^a_\mu (1 - \ap D^2) B^{a \, \mu} + \frac{1}{2} \varphi^\alpha (1- \ap D^2) \varphi^\alpha +\mathcal{V}(B,\varphi,F),
\end{equation}
and the equations of motion (EOM) take form 
\begin{equation}
    ( 1 - \ap D^2)  \varphi^\alpha = \frac{\delta \mathcal{V}}{\delta \varphi^\alpha} \equiv \mathcal{J}^\alpha \,,\quad  ( 1 - \ap D^2) B^\mu =  \frac{\delta \mathcal{V}}{\delta B^\mu} \equiv \mathcal{K}^\mu
\end{equation}
Our strategy is to use the saddle point approximation to integrate out $B$ and $\varphi$ in favor of the sources $\mathcal{J}^\alpha$ and $\mathcal{K}^\mu$, which involve powers of $F_{\mu\nu}$ as well as self interactions. The latter requires to treat the massive fields perturbatively in $\ap$. Following this procedure, the resulting Lagrangian is manifestly a function of the field strength $F_{\mu\nu}$ and its covariant derivatives. We spell out the operators contributing to the effective Lagrangian up to $\mathcal{O}(\ap^4)$. Up to order $\mathcal{O}(\ap^2)$, we match previously known expressions for the effective action \cite{Broedel:2013tta, Garozzo:2018uzj}. 

In general, the number of effective operators at each order in $\ap$ will proliferate exponentially. We are able however to write the results in terms of independent monomials of $F$ up to $\mathcal{O}(\ap^4)$, and we provide these results in a Mathematica notebook attached to this submission. Finally, we check explicitly through multiplicity six that the resulting amplitudes obey the color-kinematics relations at each of the orders in $\ap$ considered. The method can be applied systematically at arbitrary order in $\ap$ to extract the operators contributing to the zero-transcendentality sector of the open bosonic string effective action.

\section{Review}

\subsection{String amplitudes as a double copy}\label{sec:review}
The story starts with the remarkable realization that massless tree-level open superstring amplitudes can be written as the double copy (see reviews~\cite{Bern:2019prr, Bern:2022wqg, Adamo:2022dcm} and references therein) of two theories. In ref.~\cite{Mafra:2011nv}, the authors realized that open superstring amplitudes can be written in a basis of 10D Super-Yang-Mills (SYM) amplitudes which fully encode the polarization dependence of stringy amplitudes, combined with $\ap$-dependent objects originating from worldsheet integrals \cite{Broedel:2013aza, Broedel:2013tta}. These objects are interesting in their own right and can be reorganized in terms of the so-called \textit{Z-theory} amplitudes \cite{Carrasco:2016ldy, Mafra:2016mcc, Carrasco:2016ygv}. These integrals are characterized by two permutations $\pi$ and $\rho$ of $n$ labels, where the first determines the ordering of the punctures on the integration domain $D$ and the second the structure of the integrand:
\begin{equation}
    Z(\pi(1,2,\ldots,n) | \rho(1,2, \ldots, n)) = (2 \ap)^{n-3} \underset{D(\pi)}{\int} \frac{dz_1 dz_2 \ldots dz_{n-1} dz_n}{\text{vol}(SL(2, \mathbb{R}))} \frac{\prod_{i<j}^n |z_{ij}|^{ \ap s_{ij}}}{\rho(z_{12}z_{23}\ldots z_{n-1, n} z_{n, 1})},
\end{equation}
where $z_{ij} = z_i - z_j$ and $s_{ij} = 2k_i \cdot k_j $. In practice, the appearance of the inverse volume of $SL(2,\mathbb{R})$ means that the position of any three punctures $z_i, z_j, z_k$ can be fixed, in which case a Jacobian $|z_{ij} z_{jk} z_{ki}|$ should be introduced. The $Z$ integrals have been found to present a set of fascinating features: they obey monodromy relations with respect to the domain ordering $\pi$ and BCJ relations with respect to the other ordering $\rho$ \cite{Bjerrum-Bohr:2009ulz, Stieberger:2009hq}. Furthermore, in the field theory limit, they reproduce the color-stripped amplitudes of a bi-adjoint scalar theory with a cubic interaction \cite{Cachazo:2013iea}.
These properties furnished strong motivation to study and interpret $Z$ integrals as doubly-partial amplitudes for an effective theory of scalar fields. 

Having reviewed why we can talk about the $\ap$-dependent piece of stringy amplitudes as a scattering amplitude itself, we can go back to the idea that open superstring amplitudes can be thought of as a double copy between two theories and write it as a concrete statement:
\begin{equation} \label{eq:opensuper}
    A_{\text{open}}(\pi) =  \sum_{\tau, \rho \in S_{n-3}} Z(\pi|1, \tau,n,n-1) S[\tau|\rho]_1 A_{\text{SYM}}(1,\rho,n-1,n).
\end{equation}
The structure of this equation indeed reproduces that of the field-theory KLT formula \cite{Kawai:1985xq, Bjerrum-Bohr:2010pnr} where the field-theory KLT kernel $S[\tau|\rho]_1$ can be defined recursively as follows \cite{Carrasco:2016ldy} (note that $k_{ij \ldots n} = k_i + k_j + \ldots k_n$, lowercase and uppercase letters indicate single and multi-particle labels respectively, and the subscript on $S$ instructs to keep leg $i$ fixed):
\begin{equation}
S[A,j|B,j,C]_i = (k_{iB} \cdot k_j) S[A|B,C]_i, \quad \quad \quad S[\emptyset, \emptyset]_i = 1.
\end{equation}
 It is then legitimate to ask if a similar double copy form also exists for open bosonic string amplitudes. This question was considered in ref.~\cite{Huang:2016tag}, where it was determined that open bosonic string amplitudes also admit an expansion in terms of the same basis of disk integrals $Z$ considered in the superstring case, but with a crucial difference: the kinematic factors accompanying the disk integrals, which will be denoted by $B(1,2, \ldots , n|\ap)$, are now dependent on $\ap$ as well, and their leading term is given by YM amplitudes. For example, the three and four-point $B(\ldots|\ap)$ were computed \cite{Huang:2016tag} to be:
 \begin{align}
    \label{eq:bexamples}
     B(1,2,3|\ap) &= A_{\text{YM}}(1,2,3) + 4 \ap (\varepsilon_1 \cdot k_2) (\varepsilon_2 \cdot k_3) (\varepsilon_3 \cdot k_1), \\
     B(1,2,3,4|\ap)  &=  A_{\text{YM}}(1,2,3,4) + 2 \ap s_{13} \left\{ \left[ \frac{f_{12}f_{34}}{s_{12}^2 (1- \ap s_{12})} + \text{cyc}(2,3,4) \right] - \frac{4g_1 g_2 g_3 g_4}{s_{12}^2 s_{13}^2 s_{23}^2} \right\},
 \end{align}
 where $f_{ij}=(\varepsilon_i \cdot \varepsilon_j) s_{ij} - 2 (\varepsilon_i \cdot k_j) (\varepsilon_j \cdot k_i)$, $g_i = (\varepsilon_i \cdot k_{i-1}) s_{i,i+1} - (\varepsilon_i \cdot k_{i+1}) s_{i-1,i}$ and $s_{ij} = 2(k_i \cdot k_j)$.

The $\ap$ expansion of tree-level string amplitudes contains multiple zeta values (MZVs), conjecturally transcendental functions defined as a generalization of the Riemann zeta function:
\begin{equation}
    \zeta_{n_1,n_2,\ldots,n_r} = \sum_{0<k_1<k_2<\ldots<k_r}^\infty \frac{1}{k_1^{n_1} k_2^{n_2} \ldots k_r^{n_r}},
\end{equation}
where $w=n_1+n_2+\ldots n_r$ is the transcendental weight of the function. The $Z$-integrals exhibit uniform transcendentality, namely the transcendental weight of each MZV factor matches the $\ap$ order where it is found. In the open superstring amplitude expansion in terms of disk integrals and SYM amplitudes \eqref{eq:opensuper}, the MZV and $\ap$ dependence is entirely contained inside the disk integrals, hence it follows that open superstring amplitudes enjoy uniform transcendentality. This in turn implies that the $\ap$ expansion of open bosonic string amplitudes violates uniform transcendentality due to the $\ap$ dependence of the $B(\ldots|\ap)$ kinematic factors. As noted in \cite{Huang:2016tag}, at each order in $\ap$, the leading-transcendental part (MZV with highest weight) of bosonic string amplitudes is proportional to the leading term of $B(\ldots|\ap)$, which is simply $A_{\text{YM}}(\ldots)$ and hence matches the expansion for the open superstring (at the bosonic level). Therefore, the leading-transcedental piece of the amplitude expansion and the interactions that generate it can be considered universal to all open string theories. On the other hand, the zero-transcendentality piece of the open bosonic string amplitude expansion and the corresponding interactions are given by the $B(\ldots|\ap)$ expansion alone.

It would be enticing to think of the $B(\ldots|\ap)$ kinematic factors as amplitudes arising from some field theory. This idea was explored in \cite{Azevedo:2018dgo} and we will now review the intuition behind identifying such theory.

\subsection{The $(DF)^2 + {\rm YM}$ theory}
To summarize what we have learned about the $B(\ldots|\ap)$ kinematic factors, we could write them as an $\ap$ expansion
\begin{equation}
    B(1,2,\ldots,n|\ap) = A_{\text{YM}}(1,2,\ldots,n) + \sum_{k=1}^\infty \ap^k B_k(1,2\ldots,n)
\end{equation}
which reduces to YM amplitudes in the field theory limit and where we can note that the mass dimension of $B_k$ at multiplicity $n$ is $4-n-2k$. Moreover, by considering monodromy relations between open bosonic string amplitudes and selecting the piece of zero transcendentality, one can show that the $B(\ldots|\ap)$'s must satisfy BCJ relations \cite{Bern:2008qj} and the $U(1)$ decoupling identity at each order in $\ap$:
\begin{align}
    \sum_{i=2}^{n-1} k_1 \cdot k_{23 \ldots i} B_k (2,3,\ldots, i, 1, i+1, \ldots, n) &= 0, \\
    \sum_{\rho \in \text{cyclic}} B_k(1, \rho(2,3, \ldots, n)) = 0.
\end{align}
The $B_k$'s are also observed to satisfy cyclicity and Kleiss-Kuijf relations \cite{Kleiss:1988ne}, hence it is reasonable to assume they could be partial amplitudes of some gauge theory with color-kinematics duality.

Examining the high-energy limit $\ap \rightarrow \infty$, it has been noticed from explicit examples, e.g. \eqref{eq:bexamples}, that $B(\ldots|\ap)/ \ap$ contains at most double poles $s^{-2}$. Therefore we need a theory that interpolates between Yang Mills in the $\ap \rightarrow 0$ limit and a gauge theory with a four-derivative kinetic term in the $\ap \rightarrow \infty$ limit, which will be written in terms of six-dimensional operators. A unique Lagrangian for a 6D theory with color-kinematics duality was already identified in \cite{Johansson:2017srf} in the search for a theory which double copies with Yang Mills to give conformal gravity:
\begin{equation}
    \label{eq:df2lag}
    \mathcal{L}_{(DF)^2} = \frac{1}{2}(D_{\mu} F^{a\, \mu \nu})^2  +  \frac{1}{3} \,   F^3+ \frac{1}{2}(D_{\mu} \varphi^{\alpha})^2  + \frac{1}{2}  \,  C^{\alpha ab}  \varphi^{ \alpha}   F_{\mu \nu}^a F^{b\, \mu \nu }  +  \frac{1}{3!}  \, d^{\alpha \beta \gamma}   \varphi^{ \alpha}  \varphi^{ \beta} \varphi^{ \gamma}.
\end{equation}
The coupling constant $g$ was set to one and we are using the following set of conventions:
\begin{subequations} \label{eq:conventions}
\begin{align}
    F^{\mu \nu} & = \partial^\mu A^\nu - \partial^\nu A^\mu - [A^\mu, A^\nu], \\
    D^\mu F^{\nu \rho} &= \partial^\mu F^{\nu \rho} - [A^\mu, F^{\nu \rho}], \\
    D^\mu \varphi^\alpha &= \partial^\mu \varphi^\alpha - (T^a)^{\alpha \beta} A^{a \, \mu} \varphi^\beta, \\
    F^3 &= f^{abc} F_{\mu}^{a\, \nu}F_{\nu}^{b \, \rho} F_{\rho }^{c\, \mu},
\end{align}
\end{subequations}
This Lagrangian describes a gauge field $A^\mu$ and a scalar field $\varphi^\alpha$ in a real representation of the gauge group with generators $(T^a)^{\alpha \beta}$ which are antisymmetric in the last two indices. These are different from the adjoint group generators $T^a$, for which we choose conventions $[T^a, T^b] = f^{abc}T^c$ and $\text{Tr}(T^a T^b) = \delta^{ab}$. The quantities $C^{\alpha ab}$ (symmetric in last two indices) and $d^{\alpha \beta \gamma}$ (totally symmetric) are Lie algebra tensors which, in order to transform covariantly under gauge group rotations, are constrained by the following relations:
\begin{align}
    f^{abc}(T^a)^{\alpha \beta} &= (T^a)^{\alpha \gamma} (T^b)^{\gamma \beta} - (T^b)^{\alpha \gamma} (T^a)^{\gamma \beta}, \label{eq:col_1} \\
    (T^c)^{\alpha \beta} C^{\beta ab} &= f^{cae} C^{\alpha eb} + f^{cbe} C^{\alpha ea}, \label{eq:col_2} \\
    0 &= (T^a)^{\alpha \delta} d^{\delta \beta \gamma} + (T^a)^{\beta \delta} d^{\delta \gamma \alpha} + (T^a)^{\gamma \delta} d^{\delta \alpha \beta}. \label{eq:col_3}
\end{align}
A pair of extra relations exists if the amplitudes computed from the $(DF)^2$ Lagrangian are constrained to obey BCJ relations:
\begin{align}
    C^{\alpha ab} C^{\alpha cd} &= f^{ace} f^{edb} + f^{ade} f^{ecb}, \label{eq:col_4} \\
    C^{\alpha ab} d^{\alpha \beta \gamma} &= - (T^a)^{\beta \alpha} (T^b)^{\alpha \gamma} + C^{\beta ac} C^{\gamma cb} + (a \leftrightarrow b). \label{eq:col_5}
\end{align}
These five relations allow to recast the color factors of any tree-level amplitude with vectors as external states obtained from  \eqref{eq:df2lag} in terms of the structure constants $f^{abc}$ only. 

To take into account the $\ap \rightarrow 0$ limit, it is reasonable to consider augmenting this Lagrangian by a Yang-Mills term $F^2$, which would have to carry a  parameter $m^2$ with mass-dimension two. This deformation of the $(DF)^2$ theory was already studied in ref.~\cite{Johansson:2017srf} (see also \cite{ Johansson:2018ues,Azevedo:2018dgo}), where it was also noted that in order to preserve color-kinematics duality, it is necessary to introduce a mass term for the scalar field $\varphi^\alpha$. This leads to the Lagrangian
\begin{align}
\label{eq:massdefL}
{\cal L}_{(DF)^2 + {\rm YM}}&= \frac{1}{2}(D_{\mu} F^{a\, \mu \nu})^2  +  \frac{1}{3} \,   F^3+ \frac{1}{2}(D_{\mu} \varphi^{\alpha})^2  + \frac{1}{2}  \,  C^{\alpha ab}  \varphi^{ \alpha}   F_{\mu \nu}^a F^{b\, \mu \nu }  +  \frac{1}{3!}  \, d^{\alpha \beta \gamma}   \varphi^{ \alpha}  \varphi^{ \beta} \varphi^{ \gamma} \notag \\ 
 & \quad   -   \frac{1}{2} m^2 (\varphi^{\alpha})^2- \frac{1}{4} m^2 (F^a_{\mu \nu})^2.
\end{align}
This Lagrangian now interpolates between the Yang-Mills Lagrangian in the $m^2 \rightarrow \infty$ limit, and the $(DF)^2$ Lagrangian \eqref{eq:df2lag} in the  $m^2 \rightarrow 0$ limit. At this point we can identify $m^2 = - 1/ \ap$ and define a new Lagrangian \footnote{We are using an uncommon sign convention for the YM term in order to match the first few terms of the effective Lagrangian to those studied in ref.~\cite{Garozzo:2018uzj}.} $\mathcal{L} = \ap {\cal L}_{(DF)^2 + {\rm YM}}$:
\begin{align}
\label{eq:lag}
\mathcal{L} &= \frac{1}{4}(F^a_{\mu \nu})^2 + \frac{\ap}{2}(D_{\mu} F^{a\, \mu \nu})^2  +  \frac{\ap}{3} \,   F^3+ \frac{1}{2}  (\varphi^{\alpha})^2 + \frac{\ap}{2}(D_{\mu} \varphi^{\alpha})^2 + \frac{\ap}{2}  \,  C^{\alpha ab}  \varphi^{ \alpha}   F_{\mu \nu}^a F^{b\, \mu \nu }   \notag \\ 
 & \quad    +  \frac{\ap}{3!}  \, d^{\alpha \beta \gamma}   \varphi^{ \alpha}  \varphi^{ \beta} \varphi^{\gamma}.
\end{align}
The authors of ref.~\cite{Azevedo:2018dgo} have checked that the partial amplitudes for external massless gluons obtained using the Feynman rules of $\mathcal{L}$ exactly reproduce the kinematic factors $B(\ldots|\ap)$, which are known up to five points. This gives strong evidence that the field theory that appears in the double-copy expression for open-bosonic-string amplitudes is the $(DF)^2 + \rm{YM}$ theory:
\begin{equation}
    A_{\text{bos}}(\pi) =  \sum_{\tau, \rho \in S_{n-3}} Z(\pi|1, \tau ,n,n-1)  S[\tau|\rho]_1 \ap A_{(DF)^2+\rm{YM}}(1,\rho,n-1,n),
\end{equation}
and that this theory reproduces the zero-transcendentality sector of open-bosonic-string amplitudes. Our objective in this paper is to manifestly spell out the operators that generate the amplitudes of this sector. To do so we will recast the Lagrangian $\mathcal{L}$ \eqref{eq:lag} as a function of only the massless gauge field by integrating out the extra degrees of freedom, order by order in $\ap$. 

\section{The effective Lagrangian arising from the $(DF)^2+ {\rm YM}$ theory}

\subsection{Field content}
The Lagrangian \eqref{eq:lag} we considered in the previous section contains unconventional kinetic terms with four derivatives. We will not be concerned with the kinetic term of the scalar field $\varphi^\alpha$ as we are not interested in scattering this field. On the other hand, some careful considerations need to be made about the gauge field described by this Lagrangian, which we will denote from now on as $\tilde{A}^\mu$. Functions of $\tilde{A}^\mu$ are also decorated with a tilde:
\begin{align}
\label{eq:lag1}
\mathcal{L} &= \frac{1}{4}(\tilde{F}^a_{\mu \nu})^2 + \frac{\ap}{2}(\tilde{D}_{\mu} \tilde{F}^{a\, \mu \nu})^2  +  \frac{\ap}{3} \,   \tilde{F}^3+ \frac{1}{2}  (\varphi^{\alpha})^2 + \frac{\ap}{2}(\tilde{D}_{\mu} \varphi^{\alpha})^2 + \frac{\ap}{2}  \,  C^{\alpha ab}  \varphi^{ \alpha}   \tilde{F}_{\mu \nu}^a \tilde{F}^{b\, \mu \nu }   \notag \\ 
 & \quad    +  \frac{\ap}{3!}  \, d^{\alpha \beta \gamma}   \varphi^{ \alpha}  \varphi^{ \beta} \varphi^{ \gamma}.
\end{align}
Extracting the linearized equations of motion for $\tilde{A}^\mu$ from the Lagrangian \eqref{eq:lag1} and choosing gauge $\partial_\mu \tilde{A}^\mu = 0$, one finds the following fourth order equations of motion:
\begin{equation}
\Box \left( \Box - \frac{1}{\ap} \right) \tilde{A}^\mu = 0.
\end{equation}
These equations admit two plane wave solutions, one massless and one massive:
\begin{align}
\tilde{A}^\mu_0 (x) &= \varepsilon^\mu_0 e^{ik \cdot x}  \quad \text{with} \; k^2=0, \label{eq:ma1s} \\
\tilde{A}^\mu_{\ap} (x)  &= \varepsilon^\mu_{\alpha^\prime} e^{i  k \cdot x } \quad \text{with}  \; k^2  = -\frac{1}{\ap}. \label{eq:ma2s}
\end{align}
In total the gauge field $\tilde{A}^\mu$ encodes five degrees of freedom (in $D=4$ dimensions), distributed between a massless vector and a massive vector. Given that we would like to consider only gluons as external states, it is desirable to manifestly split these degrees of freedom so that a massless spin one vector $A^\mu$ enters the Lagrangian along with a massive field $B^\mu$ which carries the remaining three degrees of freedom. We will then integrate out these extra degrees of freedom, together with those of the scalar field $\varphi^\alpha$.

\subsection{Separating the Degrees of Freedom}
Our first objective is to recast the Lagrangian \eqref{eq:lag1} so that it manifestly describes a massless vector and a massive vector. This can be achieved by following the Lee-Wick procedure \cite{PhysRevD.2.1033, Grinstein:2007mp}, which reduces a four-derivative theory to one with two derivatives. The starting point is the kinematic piece of the Lagrangian (excluding that of the scalar field):
\begin{equation}
\label{eq:lkin1}
 \mathcal{L}_\text{kin.} =  \frac{1}{4} (\tilde{F}^a_{\mu \nu})^2 + \frac{\ap}{2} ( \tilde{D}_\mu \tilde{F}^{a \, \mu \nu}  )^2.
\end{equation}
First, let us encode the massive degrees of freedom in an auxiliary field defined by $B^\nu := \ap \tilde{D}_\mu \tilde{F}^{\mu \nu}$. By comparing this definition with the standard EOM in Yang-Mills one sees that the field $B^\mu$ plays the role of a conserved current and carries three degrees of freedom. Alternatively, note that the operator $ \tilde{D}_\mu \tilde{F}^{\mu \nu}$ projects out the two massless modes \eqref{eq:ma1s} leaving only the three massive modes.


With the auxiliary field, the kinetic piece of the Lagrangian can be rewritten as
\begin{equation}
\label{eq:lkin}
\mathcal{L}_\text{kin.} =  \frac{1}{4} (\tilde{F}^a_{\mu \nu})^2 -\frac{1}{2 \ap} (B^a_\mu)^2 +  B^a_\nu \tilde{D}_\mu \tilde{F}^{a \, \mu \nu}  ,
\end{equation}
which reduces to \eqref{eq:lkin1} upon inserting the equations of motion for $B^\mu$. This kinetic term only contains second derivatives but introduces non-diagonal kinetic terms between $B^\mu$ and $\tilde{A}^\mu$. In order to diagonalize the kinetic term, one can perform a change of variables $\tilde{A}^\mu = A^\mu + B^\mu$. This shift affects the entire Lagrangian generating several terms, hence it is convenient use the following definitions (where $\mathcal{O}$ is a placeholder field in a representation specified by its index):
\begin{align}
\tilde{F}^{\mu \nu} &= F^{\mu \nu} +D^\mu B^\nu - D^\nu B^\mu - [B^\mu, B^\nu], \label{eq:ftilde} \\
\tilde{D}^\mu \mathcal{O}^a &= D^\mu \mathcal{O}^a - f^{abc} B^{b \, \mu} \mathcal{O}^c, \\
\tilde{D}^\mu \mathcal{O}^\alpha &= D^\mu \mathcal{O}^\alpha -(T^a)^{\alpha \beta} B_a^\mu\mathcal{O}^\beta.
\end{align}
Also, to avoid cluttering expressions with too many structure constants, a commutator with an adjoint index is intended to be read as $[X, Y]^a = f^{abc} X^b Y^c $. The entire Lagrangian \eqref{eq:lag1} can now be rewritten as
\begin{align}
\label{eq:lag2}
 \mathcal{L} &= \frac{1}{4} (F^a_{\mu \nu})^2 - \frac{1}{2\alpha^\prime} (B^a_\mu)^2 - \frac{1}{2} (D^\mu B^a_\nu)^2 + F^a_{\mu \nu} [B^\mu, B^\nu]^a + 2 (D_\mu B^a_\nu)[B^\mu, B^\nu]^a \notag  \\
& \quad  - \frac{3}{4} ([B^\mu,B^\nu]^a)^2  + \frac{\ap}{3} \tilde{F}^3 + \frac{\ap}{2} (\tilde{D}^\mu \varphi^\alpha)^2 + \frac{1}{2} (\varphi^\alpha)^2 + \frac{\ap}{2} C^{\alpha ab} \tilde{F}^a_{\mu \nu} \tilde{F}^{b \,\mu \nu} \varphi^\alpha \notag \\
& \quad + \frac{\ap}{6} d^{\alpha \beta \gamma} \varphi^\alpha \varphi^\beta \varphi^\gamma.
\end{align}
This form of the Lagrangian manifestly contains the Yang-Mills term describing a massless vector $A^\mu$, together with a massive vector $B^\mu$ and the scalar $\varphi^\alpha$. An expression for this Lagrangian where all the $B^\mu$ dependent terms are expanded can be found in the appendix \eqref{eq:lagfull}. We are now in the position to integrate out the massive fields $B^\mu$ and $\varphi^\alpha$, in order to obtain a Lagrangian which is a function of $A^\mu$ only.

Before moving forward, let us make a remark about the $B^\mu$ equations of motion. In \eqref{eq:lkin}, $B^\mu$ was defined by the expression $B^\nu = \ap \tilde{D}_\mu \tilde{F}^{\mu \nu}$. After changing variables in the Lagrangian to obtain \eqref{eq:lag2}, this equation should still be valid upon making the same change of variables. Indeed we can compute again the equations of motion of $B^\mu$ from \eqref{eq:lag2}, and despite the fact that they will now include $\varphi^\alpha$ dependent terms, we can verify that on shell (namely after inserting the $A^\mu$ equations of motion), the two equations of motion are equivalent. This is relevant since we can use $B^\nu = \ap \tilde{D}_\mu \tilde{F}^{\mu \nu}$ to conclude that $\tilde{D} \cdot B = D \cdot B = 0$ (which was used in simplifying the Lagrangian \eqref{eq:lag2} above), and we will make use of the defining equation $B^\nu = \ap \tilde{D}_\mu \tilde{F}^{\mu \nu}$ in the following sections while working on shell.

\subsection{Massive fields expansion}
The next step is to compute the equations of motion for the fields we wish to integrate out, $B^\mu$ and $\varphi^\alpha$. For reference, variations with respect to the gauge fields are computed using formulae \eqref{eq:varying}. We obtain, for $\varphi^\alpha$ and $B^\mu$ respectively:
\begin{align}
\varphi^\alpha &= \ap D^2 \varphi^\alpha + \mathcal{J}^\alpha \notag \\
&=  \ap \Big( D^2 \varphi^\alpha -2 (T^a)^{\alpha \beta} B^a_\mu D^\mu \varphi^\beta + (T^a)^{\alpha \beta} (T^b)^{\beta \gamma} B^a_\mu B^{b \, \mu} \varphi^\gamma  \notag \\
& \quad \quad \quad - \frac{C^{\alpha ab}}{2} \tilde{F}^a_{\mu \nu} \tilde{F}^{b \, \mu \nu} -\frac{d^{\alpha \beta \gamma}}{2} \varphi^\beta \varphi^\gamma \Big), \label{eq:scalarEom1} \\
B^{a \, \lambda} &= \ap D^2 B^{a \, \lambda} + \mathcal{K}^{a \, \lambda} \notag \\
&= \ap ( D^2 B^{a \, \lambda} -2 [B_\mu,D^\mu B^\lambda]^a + [B^\mu, [B^\mu,B^\lambda]]^a - 2 C^{\alpha ab} B^{b \, \lambda} \varphi^\alpha  ) \notag \\
& \quad + (\ap)^2 ( [\tilde{D}^\lambda \tilde{F}_{\mu \nu}, \tilde{F}^{\mu \nu}]^a - (T^a)^{\alpha \beta} (\tilde{D}^\lambda \varphi^\alpha) \varphi^\beta - 2 C^{\alpha ab} \tilde{F}^{b \, \mu \lambda} (\tilde{D}_\mu \varphi^\alpha) ) \label{eq:BEom1} .
\end{align}
(see equations \eqref{eq:phiexp} and \eqref{eq:bexp} in the appendix for expressions where the tilded quantities are fully expanded). We will now consider $\varphi^\alpha$ and $B^\mu$ to be expansions in $\ap$ where each coefficient is a function of the gauge field $A^\mu$ and its derivatives. By repeatedly inserting the equations of motion into themselves, we can extract the coefficients at each order. After inspecting the above equations of motion, we can conclude that in the $\varphi^\alpha$ expansion, the lowest order term is of order $\mathcal{O}(\ap)$, while in $B^\mu$, the lowest order term is of $\mathcal{O}(\ap^2)$. Hence we will write
\begin{equation}
\varphi^\alpha = \sum_{n=1}^\infty (\ap)^n \varphi^\alpha_{(n)}   \quad \quad \text{and} \quad \quad  B^{a \, \lambda}  = \sum_{n=2}^\infty (\ap)^n B^{a \, \lambda}_{(n)}.
\end{equation}
Recall that $\tilde{F}^{\mu \nu}$, as defined in \eqref{eq:ftilde}, is a function of $B^\mu$ as well. For convenience, we will regard it as an $\ap$ expansion as well, which starts at $\mathcal{O}(\ap^2)$:
\begin{align}
    \tilde{F}^{\mu \nu} &= F^{\mu \nu} + \sum_{n=2}^\infty (\ap)^n \tilde{F}^{\mu \nu}_{(n)}.
\end{align}
The lowest order coefficients in the expansions of both $\varphi^\alpha$ and $B^\mu$ are quadratic in $F^{\mu \nu}$ and we will denote them as sources $J^\alpha$ and $K^{a \, \lambda}$ (which are the leading order terms in the sources $\mathcal{J}^\alpha$ and $\mathcal{K}^{a \, \lambda}$) for a pair of field strengths: 
\begin{align}
 \varphi^\alpha_{(1)} &= - \frac{C^{\alpha ab}}{2} F^a_{\mu \nu} F^{b \, \mu \nu} \equiv J^\alpha, \label{eq:phisource} \\ 
 B^{a \, \lambda}_{(2)} &=  [D^\lambda F_{\mu \nu},F^{\mu \nu}]^a \equiv  K^{a \, \lambda}. \label{eq:bsource}
\end{align}
 We will now recursively extract the first few coefficients of the $\varphi^\alpha$, $B^\mu$ expansions (note that $D^2 = D^\mu D_\mu$ and $D^{2k} = (D^2)^k$):
 \begin{subequations} \label{eq:fields_coefficients}
\begin{align}
    \varphi^\alpha_{(2)} &= D^2 \varphi^\alpha_{(1)}, \\
    \varphi^\alpha_{(3)} &= D^2 \varphi^\alpha_{(2)} - C^{\alpha ab} \tilde{F}^{a \, \mu \nu}_{(2)} F^b_{\mu \nu} - \frac{d^{\alpha \beta \gamma}}{2} \varphi^\beta_{(1)} \varphi^\gamma_{(1)},  \\
    \varphi^\alpha_{(4)} &= D^2  \varphi^\alpha_{(3)} -2(T^a)^{\alpha \beta} B^{a \, \mu}_{(2)} D_\mu \varphi^\beta_{(1)} - C^{\alpha ab} \tilde{F}^{a \, \mu \nu}_{(3)} F^b_{\mu \nu} - d^{\alpha \beta \gamma} \varphi^\beta_{(2)} \varphi^\gamma_{(1)}, \\
    B^{a \, \lambda}_{(3)} &= D^2 B^{a \, \lambda}_{(2)} + 2 C^{\alpha a b} F^{b \, \lambda \mu} D_\mu \varphi^\alpha_{(1)}, \\ 
        B^{a \, \lambda}_{(4)} &= D^2 B^{a \, \lambda}_{(3)} - 2 C^{\alpha ab} B^{b \, \lambda}_{(2)} \varphi^\alpha_{(1)} + [D^\lambda \tilde{F}^{\mu \nu}_{(2)}, F_{\mu \nu}]^a + [D^\lambda F_{\mu \nu}, \tilde{F}^{\mu \nu}_{(2)}]^a a \notag \\
    & \quad -2 [[B^\lambda_{(2)}, F_{\mu \nu}], F^{\mu \nu}]^a - (T^a)^{\alpha \beta} (D^\lambda \varphi^\alpha_{(1)}) \varphi^\beta_{(1)} + 2 C^{\alpha ab} F^{b \, \lambda \mu} D_\mu \varphi^\alpha_{(2)},
\end{align}
\end{subequations}
and simplify them in terms of the sources $J^\alpha$ and $K^\lambda$:
\begin{subequations}\label{eq:field_coefficients_sources}
\begin{align}
    \varphi^\alpha_{(2)} &= D^2  J^\alpha, \\
    \varphi^\alpha_{(3)} &= D^4 J^\alpha - 2 C^{\alpha ab} (D^\mu K^{a \, \nu}) F^b_{\mu \nu} - \frac{d^{\alpha \beta \gamma}}{2} J^\beta J^\gamma, \\
    \varphi^\alpha_{(4)} &= D^6 J^\alpha -2 C^{\alpha ab} (D^2 D_\mu K^a_\nu) F^{b \, \mu \nu} -4 C^{\alpha ab} (D_\rho D_\mu K_\nu^a) D^\rho F^{b \, \mu \nu}   \notag \\
    & \quad -2 C^{\alpha ab} (D_\mu K_\nu^a) D^2 F^{b \, \mu \nu} -2 C^{\alpha ab} (D_\mu D^2 K^a_\nu) F^{b \, \mu \nu} - 2 (T^a)^{\alpha \beta} K^a_\mu D^\mu J^\beta  \notag \\
    & \quad - d^{\alpha \beta \gamma} (D_\rho J^\beta) D^\rho J^\gamma -2 d^{\alpha \beta \gamma} (D^2 J^\beta) J^\gamma -4 C^{\alpha ab} C^{\beta ac} (D_\mu F^c_{\nu \rho}) (D^\rho J^\beta) F^{b \, \mu \nu}  \notag \\
    & \quad -4 C^{\alpha ab} C^{\beta ac} F^c_{\nu \rho} (D_\mu D^\rho J^\beta) F^{b \, \mu \nu}, \\
    B^{a \, \lambda}_{(3)} &= D^2 K^{a \, \lambda} + 2 C^{\alpha ab} F^{b \,  \lambda \mu} D_\mu J^\alpha, \\
    B_{(4)}^{a \, \lambda} &= D^4 K^{a \, \lambda} + 2 C^{\alpha ab} (D^2 F^{b \, \lambda \mu}) D_\mu J^\alpha + 4 C^{\alpha ab} (D^\nu F^{b \, \lambda \mu}) D^\nu D^\mu J^\alpha \notag \\
    & \quad + 2 C^{\alpha ab} F^{b \lambda \mu} D^2 D_\mu J^\alpha  +2 C^{\alpha ab} F^{b \, \lambda \mu} D_\mu D^2 J^\alpha +2 [D^\lambda D_\mu K_\nu, F^{\mu \nu}]^a \notag \\
    & \quad + 2 [D^\lambda F_{\mu \nu}, D^\mu K^\nu]^a - 2 C^{\alpha ab} K^{b \, \lambda} J^\alpha  - 2 [[K^\lambda, F_{\mu \nu}], F^{\mu \nu}]^a - (T^a)^{\alpha \beta} (D^\lambda J^\alpha) J^\beta.
\end{align}
\end{subequations}
Despite the presence of the color tensors $C^{\alpha ab}$, $d^{\alpha \beta \gamma}$ and $(T^a)^{\alpha \beta}$, we will be able to express the $\ap$ Lagrangian expansion in terms of the structure constants $f^{abc}$ only, using the relations \eqref{eq:col_1} - \eqref{eq:col_5}.

\section{Extracting $\ap$ contributions from the Lagrangian}
With an $\ap$ expansion for the $\varphi^\alpha$ and $B^\mu$ fields, we can now extract the coefficients of the Lagrangian \eqref{eq:lag2} at each order in $\ap$, where we write
\begin{equation}
    \mathcal{L} = \sum_{n=0}^\infty \ap^n \mathcal{L}_{(n)}.
\end{equation}
The first couple of coefficients are trivial to obtain. At zeroth order, the Lagrangian is just the Yang-Mills term, while at first order, it is easy to see the the only contribution comes from considering the zeroth order coefficient of $\tilde{F}^{\mu \nu}$ in the $\tilde{F}^3$ term, giving
\begin{equation}
     \mathcal{L}_{(0)} = \frac{1}{4} (F^a_{\mu \nu})^2, \quad \quad  \mathcal{L}_{(1)} = \frac{1}{3} F^3.
\end{equation}
In the following sections we will go through the manipulations needed to obtain higher order coefficients. Let us note that, since we are computing the Lagrangian on the equations of motion of $\varphi^\alpha$ and $B^\mu$, we have some freedom in the writing of the Lagrangian as we can choose to eliminate certain terms by inserting the equations of motion. For example, we could reduce the number of terms in the Lagrangian by two if we eliminate the $\ap/2(\tilde{D}^\mu \varphi^\alpha)^2$ term using the $\varphi^\alpha$ equations of motion, which allows for a more compact rewriting:
\begin{align}
\label{eq:lag3old}
 \mathcal{L} &= \frac{1}{4} (F^a_{\mu \nu})^2 - \frac{1}{2\alpha^\prime} (B^a_\mu)^2 - \frac{1}{2} (D^\mu B^a_\nu)^2 + F^a_{\mu \nu} [B^\mu, B^\nu]^a + 2 (D_\mu B^a_\nu)[B^\mu, B^\nu]^a \notag  \\
& \quad  - \frac{3}{4} ([B^\mu,B^\nu]^a)^2  + \ap \left( \frac{1}{3} \tilde{F}^3  + \frac{1}{4} C^{\alpha ab} \tilde{F}^a_{\mu \nu} \tilde{F}^{b \,\mu \nu} \varphi^\alpha - \frac{1}{12} d^{\alpha \beta \gamma} \varphi^\alpha \varphi^\beta \varphi^\gamma \right).
\end{align}
For practical purposes we will instead use the following Lagrangian
\begin{align}
\label{eq:lag3}
 \mathcal{L} &= \frac{1}{4} (F^a_{\mu \nu})^2 - \frac{1}{2\alpha^\prime} (B^a_\mu)^2 - \frac{1}{2} (D^\mu B^a_\nu)^2 + F^a_{\mu \nu} [B^\mu, B^\nu]^a + 2 (D_\mu B^a_\nu)[B^\mu, B^\nu]^a \notag  \\
& \quad  - \frac{3}{4} ([B^\mu,B^\nu]^a)^2 + \frac{1}{2} (\varphi^\alpha)^2 + \ap \left( \frac{1}{3} \tilde{F}^3  + \frac{1}{2} (D^\mu \varphi^\alpha) (D_\mu \varphi^\alpha) + (T^a)^{\alpha \beta} B^a_\mu \varphi^\alpha D^\mu \varphi^\beta  \right. \notag \\ 
& \quad \left. - \frac{1}{2} (T^a)^{\alpha \beta} (T^b)^{\beta \gamma} B^{a \, \mu} B^b_\mu \varphi^\alpha \varphi^\gamma + \frac{1}{2} C^{\alpha ab} \tilde{F}^a_{\mu \nu} \tilde{F}^{b \,\mu \nu} \varphi^\alpha + \frac{1}{6} d^{\alpha \beta \gamma} \varphi^\alpha \varphi^\beta \varphi^\gamma \right).
\end{align}
This is obtained from \eqref{eq:lag2} upon making explicit the $B^\mu$ dependence in $\tilde{D}^\mu$. At the expense of having a few more terms in the Lagrangian, this form allows for some straightforward simplifications when it comes to extracting the $\ap$ expansion coefficients.

\subsection{Lagrangian terms at $\mathcal{O}(\ap^2)$}
Inspecting \eqref{eq:lag3}, we need to account for two contributions at this order, which combine together upon using the definition of the source for $\varphi^\alpha$ \eqref{eq:phisource}:
\begin{equation}
     \mathcal{L}_{(2)} = \frac{1}{2} (\varphi^\alpha_{(1)})^2 + \frac{1}{2} C^{\alpha ab} F^a_{\mu \nu} F^{b \, \mu \nu} \varphi^\alpha_{(1)} = -\frac{1}{2} (\varphi^\alpha_{(1)})^2 = -\frac{1}{2} (J^\alpha)^2. \label{eq:prj}
\end{equation}
Once we obtain an expression for the Lagrangian coefficient at a given order in terms of the sources, we only have to expand them and manipulate the color factors into combinations of the $f^{abc}$ structure constants. For ease of notation, whenever an expression contains two $F^{\mu \nu}$ whose indices are contracted pairwise, we will denote them with the same numerical subscript. For example, $\ldots [D^\lambda F^{\mu \nu}, F^{\rho \sigma}]\ldots [F_{\mu \nu}, F_{\rho \sigma}] \ldots = \ldots [D^\lambda F_1, F_2] \ldots [F_1, F_2] \ldots $. For $\mathcal{L}_{(2)}$ we need to use the relation \eqref{eq:col_4}, $C^{\alpha ab} C^{\alpha cd} = f^{ace} f^{edb}+ f^{ade} f^{ecb}$, which leads to a single term after simplifying:
\begin{align}
\mathcal{L}_{(2)} = -\frac{1}{8} C^{\alpha ab} C^{\alpha cd} F^a_1 F^b_1 F^c_2 F^d_2 = \text{Tr} \Big( \frac{1}{4}  [F_1,F_2][F_1,F_2] \Big).
\end{align}
This agrees with the result investigated in ref.~\cite{Garozzo:2018uzj}.
It turns out this result is also evident from a tree-level Feynman diagram argument, since we can interpret \eqref{eq:prj} as a propagator Green function associated to the current $J$ (expanded in $\ap = -1/m^2 \to 0$). We outline the precise diagrams in section \ref{section:diagrams}.

\subsection{Lagrangian terms at $\mathcal{O}(\ap^3)$}
From \eqref{eq:lag3}, we extract the following terms contributing at this order:
\begin{align}
    \ \mathcal{L}_{(3)} &= -\frac{1}{2} (B^{a \, \mu}_{(2)})^2 + \varphi^\alpha_{(1)} \varphi^\alpha_{(2)} + \tilde{F}^{a \, \mu \nu}_{(2)} [{F_\nu}^\rho , F_{\rho \mu}]^a +\frac{1}{2} (D^\mu \varphi^\alpha_{(1)})^2 + \frac{1}{2} C^{\alpha ab} F^a_{\mu \nu} F^{b \, \mu \nu} \varphi^\alpha_{(2)}.
\end{align}
Note that the last term is proportional to the source and hence can be written as $-\varphi^\alpha_{(1)} \varphi^\alpha_{(2)}$, which cancels against the second term. From the structure of the Lagrangian it is clear that such cancellation always happens between the terms proportional to $\varphi^\alpha_{(1)} \varphi^\alpha_{(n-1)}$ at $\ap^n$ order.
We can further simplify this expression by examining the term proportional to $\tilde{F}^{\mu \nu}_{(2)} = D^\mu B^\nu_{(2)} - D^\nu B^\mu_{(2)}$. Integrating by parts, it can be rewritten as
\begin{equation}
\tilde{F}^{a \, \mu \nu}_{(2)} [{F_\nu}^\rho, F_{\rho \mu}]^a = 2 (D_\mu B^{a \, \nu}_{(2)})[F_{\nu \rho}, F^{\rho \mu}]^a = -2 B^{a \, \nu}_{(2)} [D_\mu F_{\nu \rho}, F^{\rho \mu}]^a + 2 B^{a \, \nu}_{(2)} [F_{\nu \rho}, D_\mu F^{\mu \rho}]^a.
\end{equation}
There is something to note about both of the terms on the right-hand side of the above equation. In the first term, we can use the fact that by the Bianchi identity, $-2[D_\mu F_{\nu \rho},F^{\rho \mu}] = [D_\nu F^{\rho \mu}, F_{\rho \mu}] = B_{\nu \, {(2)}}$.  In the second term, we generated a factor of $D_\mu F^{\mu \nu}$, which is an expansion in $\ap$. This can be seen from the definition $B^\nu = \ap \tilde{D}_\mu \tilde{F}^{\mu \nu}$ by expanding the tilded quantities:
\begin{align}
\label{eq:dfexp}
    D_\mu F^{\mu \nu} &= \frac{B^\nu}{\ap} - D^2 B^\nu +2 [B_\mu, F^{\mu \nu}] + 2 [B_\mu, D^\mu B^\nu] - [B_\mu, D^\nu B^\mu] - [B_\mu, [B^\mu, B^\nu]].
\end{align}
Since the $B^\mu$ expansion starts at second order in $\ap$, the above expansion starts at first order in $\ap$.
Assuming we are interested in Lagrangian contributions not higher than $\mathcal{O}(\ap^5)$, we can write the first few terms of the expansion
\begin{equation}
D_\mu F^{\mu \nu} = \ap B^\nu_{(2)} + (\ap)^2 \left\{ B^\nu_{(3)} - D^2 B^\nu_{(2)} + 2[B^\mu_{(2)}, {F_\mu}^\nu] \right\} + \mathcal{O}(\ap^3),
\end{equation}
and rewrite the term under scrutiny as
\begin{align}
\tilde{F}^{\mu \nu}_{(2)} [{F_\nu}^\rho, F_{\rho \mu}] &= (B^\mu_{(2)})^2 +2 \ap B^\mu_{(2)} [F_{\mu \nu}, B^\nu_{(2)}] \notag \\
& \quad +2 (\ap)^2 B^\mu_{(2)} [F_{\mu \nu},B^\nu_{(3)} - D^2 B^\nu_{(2)} + 2 [B^\rho_{(2)}, F^{\rho \nu}]]  + \mathcal{O}(\ap^3).
\end{align}
At this order we are only interested in the first term, but let us keep track of the fact that we are carrying some terms to the Lagrangian coefficients at higher order:
\begin{align}
\label{eq:extra_l4}
2  B^\mu_{(2)} [F_{\mu \nu}, B^\nu_{(2)}]  &  \in  \mathcal{L}_{(4)},  \\
\label{eq:extra_l51}
2 B^\mu_{(2)} [F_{\mu \nu},B^\nu_{(3)} - D^2 B^\nu_{(2)} + 2 [B^\rho_{(2)}, F^{\rho \nu}]]  & \in \mathcal{L}_{(5)}.
\end{align}
After these simplifications, the third order coefficient of the Lagrangian expansion in terms of the sources takes a very simple form:
\begin{equation}
     \mathcal{L}_{(3)} = \frac{1}{2} (D^\mu J^\alpha)^2 + \frac{1}{2} (K^{a \, \mu})^2.
\end{equation}
When expanding the sources, we can see that the kinematic piece of each of the above terms is identical:
\begin{align}
 \frac{1}{2} (D^\mu J^\alpha)^2 &= \frac{1}{2} C^{\alpha ab} C^{\alpha cd} (D^\mu F^a_1) F^b_1 (D_\mu F^c_2) F^d_2, \label{eq:jterm} \\
\frac{1}{2} (K^{a \, \mu})^2 &= \frac{1}{2} f^{abe}f^{ecd} (D^\mu F^a_1) F^b_1 (D_\mu F^c_2) F^d_2, \label{eq:kterm}
\end{align}
hence we only need to combine the color factors to obtain
\begin{equation} \label{eq:lag_ap3}
\mathcal{L}_{(3)}  = \text{Tr} \left( [D^\mu F_1, F_2] [D_\mu F_2, F_1] \right).
\end{equation}
Going to higher orders in $\ap$, a large number of covariant derivatives will start showing up in the Lagrangian, and introduce some freedom in writing the terms due to the possibility of integrating by parts. For example, the square of a covariant derivative acting on the field strength can be rewritten as follows using the Bianchi identity: 
\begin{equation}
    D^2 F^{\mu \nu} = D^\mu D_\rho F^{\rho \nu} - D^\nu D_\rho F^{\rho \mu} + 2 [F^{\mu \rho}, {F_\rho}^\nu].
\end{equation}
As discussed previously, by \eqref{eq:dfexp}, the terms proportional to $D_\mu F^{\mu \nu}$ are expansions in $\ap$. Terms of the form $D^2 F$ often appear in the Lagrangian coefficients due to the many powers of covariant derivatives acting on the sources. With this identity one has the freedom to move such terms to higher orders in $\ap$ and to higher multiplicities at the same order - which is not surprising given the non-uniqueness of the Lagrangian. When choosing how to write down the Lagrangian terms, our guiding principle will be to attempt to write the kinematic part (monomials in $F$ or $DF$) of different terms into the same tensor structures. For example, in the $\mathcal{O}(\ap^3)$ case above, we chose to write the $J$-dependent term \eqref{eq:jterm} by having each of the two derivatives acting on one of the sources, which has the same structure as the $K$-dependent term \eqref{eq:kterm}.

\subsection{Lagrangian terms at $\mathcal{O}(\ap^4)$}
At this order we have to consider the following contributions:
\begin{align}
     \mathcal{L}_{(4)} &= - B^{a \, \mu}_{(2)} B^a_{(3) \, \mu} - \frac{1}{2} (D^\mu B^{a \, \nu}_{(2)})^2 + F^a_{\mu \nu}[B^\mu_{(2)}, B^\nu_{(2)}] + \varphi^\alpha_{(1)} \varphi^\alpha_{(3)} + \frac{1}{2} \varphi^\alpha_{(2)} \varphi^\alpha_{(2)}  \notag \\
    & \quad + \tilde{F}^{a \, \mu \nu}_{(3)} [F_{\nu \rho}, {F^\rho}_\mu]^a + (D^\mu \varphi^\alpha_{(1)}) (D_\mu \varphi^\alpha_{(2)}) + \frac{1}{2} C^{\alpha ab} F^a_{\mu \nu} F^{b \, \mu \nu} \varphi^\alpha_{(3)} + \frac{1}{2} C^{\alpha ab} \tilde{F}^{a \, \mu \nu}_{(2)} F^b_{\mu \nu} \varphi^\alpha_{(1)} \notag \\
    & \quad + \frac{1}{6}d^{\alpha \beta \gamma} \varphi^\alpha_{(1)} \varphi^\beta_{(1)} \varphi^\gamma_{(1)} - 2 F^a_{\mu \nu} [B^\mu_{(2)}, B^\nu_{(2)}]^a.
\end{align}
Notice that the last term was not generated from the Lagrangian \eqref{eq:lag3} but rather carried from the computation at $\mathcal{O}(\ap^3)$ \eqref{eq:extra_l4}.
As before, we can inspect the term proportional to $\tilde{F}^{\mu \nu}_{(3)}$ and rewrite it as
\begin{equation}
    \tilde{F}^{\mu \nu}_{(3)} [F_{\nu \rho}, {F^\rho}_\mu] = 2 D^\mu B^\nu_{(3)} [F_{\nu \rho}, {F^\rho}_\mu] = B^\mu_{(3)} B_{(2) \, \mu} + 2 \ap B^\mu_{(3)} [F_{\mu \nu}, B^\nu_{(2)}] + \mathcal{O}(\ap^2). 
\end{equation}
Again, we only consider the first term and add the second to the collection of contributions pushed to the next order:
\begin{equation} \label{eq:extra_l52}
     2 B^\mu_{(3)} [F_{\mu \nu}, B^\nu_{(2)}] \in \mathcal{L}_{(5)}.
\end{equation}
Simplifying the remaining terms one obtains the Lagrangian as a function of the sources at $\mathcal{O}(\ap^4)$:
\begin{align}
     \mathcal{L}_{(4)} &= -\frac{1}{2} J^\alpha D^4 J^\alpha - \frac{1}{2} (D^\mu K^{a \, \nu})^2  - F^a_{\mu \nu} [K^\mu, K^\nu]^a +2 C^{\alpha ab} (D^\mu K^{a \, \nu} ) F^b_{\mu \nu} J^\alpha \notag \\
    & \quad + \frac{1}{6} d^{\alpha \beta \gamma} J^\alpha J^\beta J^\gamma.
    \label{eq:lag4}
\end{align}
The first two terms have schematic form $D^4 F^4$, the next two  $D^2 F^5$ and the last as $F^6$. Let us evaluate them starting from the first two, which begin contributing at multiplicity four. Ideally, we would recast them such that their kinematic part is similar so that we just need to combine and manipulate their color factors is a convenient basis. Since schematically $K \sim D F^2$ and $J \sim F^2$, a similar tensor structure for both terms can be achieved if both are integrated by parts until there are no squares of covariant derivatives. Explicitly, we can write:
\begin{align}
 -\frac{1}{2} (D^\mu K^{a \, \nu})^2 = -\frac{1}{2} f^{abe}f^{ecd} & \left\{ (D^\mu D^\nu F^a_1)F^b_1 (D_\mu D_\nu F^c_2) F^d_2 - 2 (D^\mu D^\nu F^a_1) F^b_1 (D_\mu F^c_2) (D_\nu F^d_2) \right. \notag \\
& \quad \left. + (D^\mu F^a_1) (D^\nu F^b_1) (D_\mu F^c_2) (D_\nu F^d_2) \right\}. \label{eq:dksquared}
\end{align}

\begin{align}
- \frac{1}{2} J^\alpha (D^4 J^\alpha) &=  \frac{1}{2} (D^\mu J^\alpha) (D_\mu D_\nu D^\nu J^\alpha) \notag \\
&= \frac{1}{2} (D^\mu J^\alpha)(D^\nu D_\mu D_\nu J^\alpha) + \frac{1}{2} (D^\mu J^\alpha) [D_\mu, D_\nu] (D^\nu J^\alpha) \notag  \\
&= -\frac{1}{2} (D^\mu D^\nu J^\alpha) (D_\nu D_\mu J^\alpha) - \frac{1}{2} (T^a)^{\alpha \beta} F_{\mu \nu}^a (D^\mu J^\alpha) (D^\nu J^\beta) \label{eq:d4j2}
\end{align}
where in the last line we used $[D^\mu , D^\nu] \mathcal{O}^\alpha = -(T^a)^{\alpha \beta} F^{\mu \nu,a} \mathcal{O}^\beta$, which  generates a term contributing to the $D^2 F^5$ piece of $\mathcal{L}_{(4)}$. We will now focus on the first term of \eqref{eq:d4j2}, whose kinematic part has some terms in common with \eqref{eq:dksquared}:
\begin{align} 
-\frac{1}{2} (D^\mu D^\nu J^\alpha)(D_\nu D_\mu J^\alpha) &= -\frac{1}{2} C^{\alpha ab} C^{\alpha cd} \left\{ (D^\mu D^\nu F^a_1)F^b_1 (D_\nu D_\mu F^c_2) F^d_2  \right. \notag \\
& \quad  + 2 (D^\mu D^\nu F^a_1) F^b_1 (D_\mu F^c_2) (D_\nu F^d_2) \notag \\
& \quad \left. + (D^\mu F^a_1) (D^\nu F^b_1) (D_\mu F^c_2) (D_\nu F^d_2) \right\}.
\label{eq:t1c}
\end{align}
Note that by rewriting $D_\nu D_\mu F^c_2 = D_\mu D_\nu F_2^c + [F_{\mu \nu}, F_2]^c$ in the first line of \eqref{eq:t1c}, the two expressions contributing to the $D^4F^4$ piece will present the same tensor structure. This generates a term which starts contributing at six points and will be evaluated later.
Simplifying the color factors, we can combine the $D^4 F^4$ operators at this order as
\begin{align}
\mathcal{L}_{(4)} \vert_{D^4 F^4} &= \text{Tr} \left( [D^\mu D^\nu F_1, F_2] [F_1, D_\mu D_\nu F_2] + [D^\mu F_1, D^\nu F_2] [D_\nu F_1, D_\mu F_2] \right. \notag \\
& \quad \left. + 2 [D^\mu D^\nu F_1, D_\mu F_2] [F_1, D_\nu F_2] \right). \label{eq:lag_ap4_f4}
\end{align}
Moving on to the $D^2 F^5$ terms,  we start from the one that was generated in \eqref{eq:d4j2}:
\begin{align}
-\frac{1}{2} (T^e)^{\alpha \beta} F_{\mu \nu}^e (D^\mu J^\alpha) (D^\nu J^\beta) &= -\frac{1}{2} (T^e)^{\alpha \beta} C^{\alpha ab} C^{\beta cd} (D^\mu F^a_1)F^b_1 (D^\nu F^c_2) F^d_2 F_{\mu \nu}^e, \label{eq:term5}
\end{align}
for which we need to use the relation \eqref{eq:col_3}, $(T^c)^{\alpha \beta} C^{\beta ab} = f^{cae} C^{\alpha eb} +f^{cbe} C^{\alpha ea}$. The kinematic part of \eqref{eq:term5} combines with that of the following term:
\begin{align}
-F_{\mu \nu} [K^\mu , K^\nu] = -f^{e h g} f^{h ab} f^{g cd} (D^\mu F^a_1) F^b_1 (D^\nu F^c_2) F^d_2 F_{\mu \nu}^e,
\end{align}
while the last $D^2 F^5$ operator gives an independent kinematic structure plus a term which starts contributing at six points due to the antisymmetry of the field strength:
\begin{align}
2 C^{\alpha ab} (D^\mu K^{\nu,a}) F_{\mu \nu}^b J^\alpha &=  C^{\alpha he} C^{\alpha cd} f^{hab} \left\{ (D^\mu F^a_1) (D^\nu F^b_1) F^c_2 F^d_2 F_{\mu \nu}^e  \right. \notag \\
& \quad \left. - (D^\mu D^\nu F^a_1)F^b_1 F^c_2 F^d_2 F_{\mu \nu}^e  \right\} .
\label{eq:t4a}
\end{align}
Combining the color factors in a convenient manner, we find that the Lagrangian contains the following operators which start contributing at five points:
\begin{align}
\mathcal{L}_{(4)} \vert_{D^2 F^5} &= \text{Tr} \Big( \frac{1}{2} [D^\mu F_1, F_1] [F_{\mu \nu}, [D^\nu F_2, F_2]] + [D^\mu F_1, D^\nu F_2] [F_2, [F_{\mu \nu}, F_1]] \notag \\
& \quad  + [D^\mu F_1, F_{\mu \nu}] [D^\nu F_2, [F_1, F_2]] + 2 [D^\mu F_1, D^\nu F_1] [F_2, [F_2, F_{\mu \nu}]] \Big).
\end{align}
At last, we consider the $F^6$ term coming from \eqref{eq:lag4}:
\begin{align}
\frac{1}{6} d^{\alpha \beta \gamma} J^\alpha J^\beta J^\gamma &= -\frac{1}{48} d^{\alpha \beta \gamma} C^{\alpha ab} C^{\beta cd} C^{\gamma ef} F^a_1 F^b_1 F^c_2 F^d_2 F^e_3 F^f_3 \notag \\
&=   -\frac{1}{3} \text{Tr} \Big( [F_1,F_2][[F_1,F_3],[F_2,F_3]] \Big),
\end{align}
where to simplify the color factors we have used the following identity, which comes from a combination of the relations \eqref{eq:col_2} and \eqref{eq:col_5}:
\begin{align}
d^{\alpha \beta \gamma} C^{\alpha ab} C^{\beta cd} &= C^{\gamma ae} C^{\delta eb} C^{\delta cd} + C^{\gamma be} C^{\delta ea} C^{\delta cd} +C^{\gamma ce} C^{\delta ed} C^{\delta ab} + C^{\gamma de} C^{\delta ec} C^{\delta ab} \notag \\
& \quad +2 C^{\gamma eg} (f^{eac}f^{gbd} + f^{ead}f^{gbc}).
\end{align}
While computing the $D^4 F^4$ and $D^2 F^5$ pieces of the Lagrangian, we identified two more terms proportional to $F^6$. From \eqref{eq:t1c},
\begin{align}
    -\frac{1}{2} C^{\alpha ab} C^{\alpha hd} f^{hec} (D^\mu D^\nu F_1^a) F_1^b F_2^c F_2^d F_{\mu \nu}^e & = \frac{1}{4} C^{\alpha gb} C^{\alpha hd} f^{h ec} f^{gfa} F^a_1 F^b_1 F^c_2 F^d_2 F^e_3 F^f_3 \notag \\
    & =  -\frac{1}{4} \text{Tr} \Big( [F_1, F_2] [ [ F_1, F_3], [F_2, F_3]] \notag \\
    & \quad \quad \quad \quad + [F_1, F_2] [ F_3, [F_2, [F_1, F_3]]] \Big),
\end{align}
and from \eqref{eq:t4a},
\begin{align}
    -C^{\alpha he} C^{\alpha cd} f^{hab} (D^\mu D^\nu F_1^a) F_1^b F_2^c F_2^d F_{\mu \nu}^e & = \frac{1}{2} C^{\alpha he} C^{\alpha cd} f^{hgb} f^{gfa} F^a_1 F^b_1 F^c_2 F^d_2 F^e_3 F^f_3 \notag \\
    & =  \text{Tr} \Big( [F_1, F_2] [ [ F_1, F_3], [F_2, F_3]] \notag \\
    &  \quad \quad \quad -  [F_1, F_2] [ F_3, [F_2, [F_1, F_3]]] \Big).
\end{align}
Combining the $F^6$ operators, we obtain
\begin{equation}
    \mathcal{L}_{(4)} \vert_{F^6} = \text{Tr} \Big( \frac{5}{12} [F_1, F_2] [ [ F_1, F_3], [F_2, F_3]] - \frac{5}{4} [F_1, F_2] [ F_3, [F_2, [F_1, F_3]]] \Big).
\end{equation}
The full Lagrangian at order $\ap^4$ is then
\begin{align}
    \mathcal{L}_{(4)} &= \text{Tr} \Big( [D^\mu D^\nu F_1, F_2] [F_1, D_\mu D_\nu F_2] + [D^\mu F_1, D^\nu F_2] [D_\nu F_1, D_\mu F_2]  \notag \\
    & \quad \quad \quad + 2 [D^\mu D^\nu F_1, D_\mu F_2] [F_1, D_\nu F_2] + \frac{1}{2} [D^\mu F_1, F_1] [F_{\mu \nu}, [D^\nu F_2, F_2]]  \notag \\
    & \quad \quad \quad + [D^\mu F_1, D^\nu F_2] [F_2, [F_{\mu \nu}, F_1]] + [D^\mu F_1, F_{\mu \nu}] [D^\nu F_2, [F_1, F_2]] \notag \\
    & \quad \quad \quad + 2 [D^\mu F_1, D^\nu F_1] [F_2, [F_2, F_{\mu \nu}]] + \frac{5}{12} [F_1, F_2] [ [ F_1, F_3], [F_2, F_3]] \notag \\
    & \quad \quad \quad - \frac{5}{4} [F_1, F_2] [ F_3, [F_2, [F_1, F_3]]] \Big).
\end{align}
The validity of this expression was tested by computing the amplitudes through multiplicity six and verifying that they obey BCJ relations at each order in $\ap$ up to $\mathcal{O}(\ap^4)$. The amplitudes were also checked to match the known $B(\ldots|\ap)$ factors, which are known to all orders up to multiplicity five \cite{Huang:2016tag}. \footnote{To compute the scattering amplitudes, we find it convenient to extract the equations of motion for the gauge field $A^\mu$ from the Lagrangian and use the perturbiner expansion \cite{Rosly:1996vr, Rosly:1997ap, Selivanov:1997aq, Selivanov:1999as} to write a recursion for the rank $n-1$ Berends-Giele currents \cite{Berends:1987me}, which are then used to compute $n$-point amplitudes. See refs.~\cite{Lee:2015upy, Mafra:2015vca, Mizera:2018jbh} for more details on this method.}

\subsection{Higher order in $\ap$}
We expect that applying the same procedure to higher orders in $\ap$ will continue yielding the operators appearing in the zero-transcendentality sector of the effective Lagrangian for the open bosonic string. As expected, the number of terms contributing at each order considerably increases. For example, after simplifying the terms obtained from \eqref{eq:lag3} and including those we carried from the previous order computations \eqref{eq:extra_l51} and \eqref{eq:extra_l52}, the $\mathcal{O}(\ap^5)$ Lagrangian coefficient in terms of the sources is given by
\begin{align}
    \mathcal{L}_{(5)} &= -\frac{1}{2} J^\alpha D^6 J^\alpha + \frac{1}{2} K^{a \, \mu} D^4 K^a_\mu + 2 C^{\alpha ab} (D^\mu K^{a, \nu}) F^b_{\mu \nu} D^2 J^\alpha - 2 C^{\alpha ab} (D^2 K^{a \, \nu}) F^b_{\mu \nu} D^\mu J^\alpha  \notag \\
    & \quad + 2 F^a_{\mu \nu} [D^\nu K^\rho, D_\rho K^\mu]^a - F^a_{\mu \nu} [D^\rho K^\nu, D_\rho K^\mu]^a + \frac{1}{2} \left([F^{\mu \nu}, K^\rho]^a\right)^2 +(T^a)^{\alpha \beta} K^a_\mu J^{\alpha} D^\mu J^\beta \notag \\
    & \quad  + 4[K_\mu, F^{\mu \nu}]^a [K^\rho, F_{\rho \nu}]^a - [K^\mu, K_\nu]^a [F^{\nu \rho}, F_{\rho \mu}]^a -4 C^{\alpha ab} [F^{\mu \nu}, K_\mu]^a F^b_{\nu \rho} D^\rho J^\alpha \notag \\
    & \quad  + 2 C^{\alpha ab} C^{\beta ac} F^{b \, \mu \nu} F^c_{\mu \rho} (D_\nu J^\alpha) D^\rho J^\beta + \frac{1}{3} d^{\alpha \beta \gamma} (D^2 J^\alpha) J^\beta J^\gamma,
\end{align}
where for completeness we report again the definitions for the sources:
\begin{align}
 J^\alpha &= - \frac{C^{\alpha ab}}{2} F^a_{\mu \nu} F^{b \, \mu \nu}, \\ 
 K^{a \, \lambda} &=  [D^\lambda F_{\mu \nu},F^{\mu \nu}]^a.
\end{align}
After expanding the sources and manipulating the color factors into strings of structure constants, the above Lagrangian will have the following schematic structure:
\begin{equation}
    \mathcal{L}_{(5)} = D^6 F^4 + D^4 F^5 + D^2 F^6,
\end{equation}
but we can expect the presence of $F^7$ operators, since some terms might be pushed to higher multiplicity due to the antisymmetry of the field strength. By dimensional analysis, the Lagrangian at $\mathcal{O}(\ap^n)$ for $n \geq 2$ is indeed expected to contain the following operators:
\begin{equation}
\mathcal{L}_{(n)} \sim \sum_{m=4}^{n+2} D^{2(n-m+2)} F^m.
\end{equation}

From a computational standpoint, it is simple to use the expressions in the appendix for the Lagrangian and the equations of motion of the fields $\varphi^\alpha$ and $B^\mu$ to set up a recursion for their coefficients at arbitrary order in $\ap$, in terms of the sources. There is, however, a lot of freedom involved in the process of recasting the Lagrangian as a trace. It would be interesting to have a well-defined procedure for carrying out this step which lands on a simple expression for the Lagrangian coefficients, or to explore possible different rewritings of the Lagrangian which would allow for a more efficient extraction of the coefficients at each order in $\ap$. 

\section{Diagrammatic expansion} \label{section:diagrams}
From the above considerations we see that it would be desirable to implement an organizing principle for the effective action at each $\ap$ order. With this in mind, let us here re-examine the Lagrangian construction from a diagrammatic viewpoint. Using this, we will show how to reconstruct the operators proportional to $F^4$ at all orders in $\ap$, and leave higher-point extensions to future work.

Recall that the fields we want to integrate out from the Lagrangian, $\varphi^\alpha$ and $B^\mu$, have sources $J^\alpha$ and $K^\mu$ respectively which are quadratic in the field strength $F$. If we are interested in operators of order not higher than $F^4$, we can neglect all the other couplings and consider linearized solutions to the equations of motion, which only depend on the source terms:
\begin{align}
    \varphi^\alpha &\sim \ap G J^\alpha, \label{eq:solphi} \\
    B^\lambda &\sim \ap^2 G K^\lambda, \quad \text{with} \quad  G= \frac{1}{1- \ap D^2}.
    \label{eq:solb}
\end{align}
Here $G$ is a propagator, and in the low energy limit we can regard it as a geometric series in the local operator $\ap D^2$, thus generating an expansion in $\ap$. In the Lagrangian \eqref{eq:lag1} we only have to consider those terms that are at most quadratic in the fields $\varphi^\alpha$ and $B^\mu$ (or quartic in $F$):
\begin{align}
    \mathcal{L}|_{F^4} &= \frac{\varphi^2}{2} - \frac{\ap}{2} \varphi^\alpha D^2 \varphi^\alpha - \ap J^\alpha \varphi^\alpha -\frac{B^2}{2 \ap} + \frac{1}{2} B^\mu D^2 B_\mu + \ap K^\mu B_\mu \\
    &= \frac{1}{2} \varphi^\alpha G^{-1} \varphi^\alpha - \ap J^\alpha \varphi^\alpha - \frac{1}{2 \ap} B^\mu G^{-1} B_\mu + \ap K^\mu B_\mu.
\end{align}
These are simply the kinetic and source terms for the fields we wish to integrate out, which we wrote in terms of the inverse propagator $G^{-1} = (1- \ap D^2)$ in the second line. Note that the last term, which does not appear explicitly in the Lagrangian \eqref{eq:lag1}, was obtained by integrating by parts the $F^4$ piece of the $\tilde{F}^3$ operator (this can be seen manifestly by looking at the fully expanded expression in the appendix \eqref{eq:lagfull}). Inserting the solutions \eqref{eq:solphi} and \eqref{eq:solb}, one obtains the non-local expression
\begin{equation} \label{eq:nonloc}
    \mathcal{L}|_{F^4} = -\frac{\ap^2}{2} J^\alpha G J^\alpha + \frac{\ap^3}{2} K^\mu G K_\mu,
\end{equation}
We can interpret each term in this expression as a graph where two sources are connected by a propagator $G$, whose presence is denoted by a pair of vertical lines:
 \begin{equation}
    \mathcal{L}|_{F^4} =
    \begin{tikzpicture}[line width=0.2mm, baseline=-3pt]
        \draw(0,0)node{$\bullet$}node[left]{$J$}--(1,0)node{$\bullet$}node[right]{$J$};
        \node at (0.5,0) {$\parallel$};
    \end{tikzpicture}
+  
\begin{tikzpicture}[line width=0.2mm, baseline=-3pt]
\draw[snake it](0,0)node{$\bullet$}node[left]{$K$}--(1,0)node{$\bullet$}node[right]{$K$};
\node at (0.5,0) {$\parallel$};
\end{tikzpicture}
.
\end{equation}
Solid lines carry the degrees of freedom of the colored scalar, while wavy lines are used for fields with adjoint indices. For each diagram, we associate a factor of $\ap$ to each source, a symmetry factor of 1/2 and a normalization factor:
\begin{align}
    \begin{tikzpicture}[line width=0.2mm, baseline=-3pt]
        \draw(0,0)node{$\bullet$}node[left]{$J$}--(1,0)node{$\bullet$}node[right]{$J$};
        \node at (0.5,0) {$\parallel$};
    \end{tikzpicture}
&= -\frac{1}{2} (\ap J^\alpha) G (\ap J^\alpha), \\
    \begin{tikzpicture}[line width=0.2mm, baseline=-3pt]
        \draw(0,0)[snake it]node{$\bullet$}node[left]{$K$}--(1,0)node{$\bullet$}node[right]{$K$};
        \node at (0.5,0) {$\parallel$};
    \end{tikzpicture}
&= \frac{\ap}{2} (\ap K^\mu) G (\ap K_\mu).
\end{align}
The propagator $G$ is to be considered as an expansion in $\ap$ whose operators can act on either side of the line:
\begin{equation}
        \begin{tikzpicture}[line width=0.2mm, baseline=-3pt]
\draw(0,0)--(1,0);
\node at (0.5,0) {$\parallel$};
\end{tikzpicture}
= \sum_{n=0}^\infty (\ap D^2)^n.
\end{equation}
To extract the contributions at each order in $\ap$ we have to select the appropriate term from the $G = 1 + \ap D^2 + \ap^2 D^4 + \ldots$ expansion. In the following diagrams we will explicitly indicate the selected operators above the propagator line. The final step consists in attaching two gluon lines to the sources with the appropriate numerical factors. The color factors remain implicit in the vertices, namely
\begin{equation}
    \begin{tikzpicture}[line width=0.2mm, baseline=-3pt, scale=0.9]
        \draw(0,0)--(1,0)node[right]{$\alpha$};
        \draw[snake it](0,0)--(-0.75,-0.75)node[left]{$a$};
        \draw[snake it](0,0)--(-0.75,0.75)node[left]{$b$};
    \end{tikzpicture}
    \sim C^{\alpha ab}
    \quad \quad \text{and} \quad \quad 
    \begin{tikzpicture}[line width=0.2mm, baseline=-3pt, scale=0.9]
        \draw[snake it](0,0)--(1,0)node[right]{$c$};
        \draw[snake it](0,0)--(-0.75,-0.75)node[left]{$a$};
        \draw[snake it](0,0)--(-0.75,0.75)node[left]{$b$};
    \end{tikzpicture}
    \sim f^{abc}.
\end{equation}
For example, at $\mathcal{O}(\ap^2)$, where only the $J$ graph contributes, we can perform the computation as follows:
\begin{align}
\mathcal{L}_{(2)}|_{F^4} &= 
\begin{tikzpicture}[line width=0.2mm, baseline=-3pt]
    \draw(0,0)node{$\bullet$}node[left]{$J$}--(1,0)node{$\bullet$}node[right]{$J$};
\end{tikzpicture}
 = -\frac{1}{2} \times \left(-\frac{1}{2}\right)^2
    \begin{tikzpicture}[line width=0.2mm, baseline=-3pt]
        \draw(0,0)--(1,0);
\draw[snake it](0,0)--(-0.75, 0.75)node{$\bullet$}node[left]{$F_1$};
\draw[snake it](0,0)--(-0.75, -0.75)node{$\bullet$}node[left]{$F_1$};
\draw[snake it](1,0)--(1.75, 0.75)node{$\bullet$}node[right]{$F_2$};
\draw[snake it](1,0)--(1.75, -0.75)node{$\bullet$}node[right]{$F_2$};
    \end{tikzpicture} \\
&= -\frac{1}{2} \times \left(-\frac{1}{2}\right)^2 \times 2
    \begin{tikzpicture}[line width=0.2mm, baseline=-3pt]
        \draw[snake it](-0.75, 0.75)node{$\bullet$}node[left]{$F_2$}--(1.75, 0.75)node{$\bullet$}node[right]{$F_1$};
        \draw[snake it](0.5,0.75)--(0.5,-0.75);
        \draw[snake it](-0.75,-0.75)node{$\bullet$}node[left]{$F_2$}--(1.75,-0.75)node{$\bullet$}node[right]{$F_1$};
    \end{tikzpicture} \notag \\ 
    &= \frac{1}{4} \text{Tr}\left([F_1, F_2] [F_1, F_2]\right).
\end{align}
The factor of 2 in the last diagram is explained by considering that upon expanding the scalar line with the color factor relation \eqref{eq:col_4}, the gluons in each pair are indistinguishable. This relation can be represented diagrammatically as
\begin{equation}
    \begin{tikzpicture}[line width=0.2mm, baseline=-3pt]
        \draw(0,0)--(1,0);
\draw[snake it](0,0)--(-0.75, 0.75)node[left]{$b$};
\draw[snake it](0,0)--(-0.75, -0.75)node[left]{$a$};
\draw[snake it](1,0)--(1.75, 0.75)node[right]{$c$};
\draw[snake it](1,0)--(1.75, -0.75)node[right]{$d$};
    \end{tikzpicture}
    = 
    \begin{tikzpicture}[line width=0.2mm, baseline=-3pt]
        \draw[snake it](-0.75, 0.75)node[left]{$c$}--(1.75, 0.75)node[right]{$b$};
        \draw[snake it](0.5,0.75)--(0.5,-0.75);
        \draw[snake it](-0.75,-0.75)node[left]{$d$}--(1.75,-0.75)node[right]{$a$};
    \end{tikzpicture} 
    +
    \begin{tikzpicture}[line width=0.2mm, baseline=-3pt]
        \draw[snake it](-0.75, 0.75)node[left]{$d$}--(1.75, 0.75)node[right]{$b$};
        \draw[snake it](0.5,0.75)--(0.5,-0.75);
        \draw[snake it](-0.75,-0.75)node[left]{$c$}--(1.75,-0.75)node[right]{$a$};
    \end{tikzpicture} 
    .
\end{equation}
 Starting from $\mathcal{O}(\ap^3)$, the Lagrangian receives contributions from both diagrams:
\begin{align}
    \mathcal{L}_{(3)}|_{F^4} &= 
    \begin{tikzpicture}[line width=0.2mm, baseline=-3pt]
    \draw(0,0)node{$\bullet$}node[left]{$J$}--(1,0)node{$\bullet$}node[right]{$J$};
    \node at (0.55,0.25) {$D^2$};
\end{tikzpicture}
+
    \begin{tikzpicture}[line width=0.2mm, baseline=-3pt]
    \draw(0,0)[snake it]node{$\bullet$}node[left]{$K$}--(1,0)node{$\bullet$}node[right]{$K$};
\end{tikzpicture} \notag \\
&= - \left(-\frac{1}{2} \right) \times \left( -\frac{1}{2} \right)^2 \times 2^2
\begin{tikzpicture}[line width=0.2mm, baseline=-3pt, scale=0.8]
        \draw(0,0)--(1,0);
\draw[snake it](0,0)--(-0.75, 0.75)node{$\bullet$}node[left]{$F_1$};
\draw[snake it](0,0)--(-0.75, -0.75)node{$\bullet$}node[left]{$D^\mu F_1$};
\draw[snake it](1,0)--(1.75, 0.75)node{$\bullet$}node[right]{$D_\mu F_2$};
\draw[snake it](1,0)--(1.75, -0.75)node{$\bullet$}node[right]{$F_2$};
    \end{tikzpicture}
    + \frac{1}{2} 
    \begin{tikzpicture}[line width=0.2mm, baseline=-3pt, scale=0.8]
        \draw[snake it](-0.40, 0.75)node{$\bullet$}node[left]{$F_1$}--(1.40, 0.75)node{$\bullet$}node[right]{$D^\mu F_2$};
        \draw[snake it](0.5,0.75)--(0.5,-0.75);
        \draw[snake it](-0.40,-0.75)node{$\bullet$}node[left]{$D_\mu F_1$}--(1.40,-0.75)node{$\bullet$}node[right]{$F_2$};
    \end{tikzpicture} \notag \\
    &= \frac{1}{2} \left(
    \begin{tikzpicture}[line width=0.2mm, baseline=-3pt, scale=0.8]
        \draw[snake it](-0.40, 0.75)node{$\bullet$}node[left]{$D^\mu F_2$}--(1.40, 0.75)node{$\bullet$}node[right]{$F_1$};
        \draw[snake it](0.5,0.75)--(0.5,-0.75);
        \draw[snake it](-0.40,-0.75)node{$\bullet$}node[left]{$F_2$}--(1.40,-0.75)node{$\bullet$}node[right]{$D_\mu F_1$};
    \end{tikzpicture}
    +
    \begin{tikzpicture}[line width=0.2mm, baseline=-3pt, scale=0.8]
        \draw[snake it](-0.40, 0.75)node{$\bullet$}node[left]{$F_2$}--(1.40, 0.75)node{$\bullet$}node[right]{$F_1$};
        \draw[snake it](0.5,0.75)--(0.5,-0.75);
        \draw[snake it](-0.40,-0.75)node{$\bullet$}node[left]{$D^\mu F_2$}--(1.40,-0.75)node{$\bullet$}node[right]{$D_\mu F_1$};
    \end{tikzpicture}
     + 
 \begin{tikzpicture}[line width=0.2mm, baseline=-3pt, scale=0.8]
        \draw[snake it](-0.40, 0.75)node{$\bullet$}node[left]{$F_1$}--(1.40, 0.75)node{$\bullet$}node[right]{$D^\mu F_2$};
        \draw[snake it](0.5,0.75)--(0.5,-0.75);
        \draw[snake it](-0.40,-0.75)node{$\bullet$}node[left]{$D_\mu F_1$}--(1.40,-0.75)node{$\bullet$}node[right]{$F_2$};
    \end{tikzpicture} \right) \notag \\
    &=
    \begin{tikzpicture}[line width=0.2mm, baseline=-3pt, scale=0.8]
        \draw[snake it](-0.40, 0.75)node{$\bullet$}node[left]{$D^\mu F_2$}--(1.40, 0.75)node{$\bullet$}node[right]{$F_1$};
        \draw[snake it](0.5,0.75)--(0.5,-0.75);
        \draw[snake it](-0.40,-0.75)node{$\bullet$}node[left]{$F_2$}--(1.40,-0.75)node{$\bullet$}node[right]{$D_\mu F_1$};
    \end{tikzpicture}
    = \text{Tr}\left([D^\mu F_1, F_2][D_\mu F_2, F_1] \right).
\end{align}
We have integrated by parts and used the Jacobi identity to reproduce the Lagrangian coefficients in the same form as in the previous sections. 

When considering higher order in $F$ contributions, we expect that studying the Lagrangian in terms of the sources $J$ and $K$ and the propagators $G$, could offer a more efficient method of computing the operators at each order in $\ap$. As a final remark, the $F^4$ Lagrangian expression \eqref{eq:nonloc} can be simplified into a single term \eqref{eq:lag_f4_full} which gives the $F^4$ operators at all orders in $\ap$.

\section{Conclusions}
We have illustrated a method to obtain the operators of the zero-transcendentality sector of the effective Lagrangian for the open bosonic string, at arbitrary order in $\ap$. We have carried out explicit computation of the operators up to $\mathcal{O}(\ap^4)$, and checked that the resulting amplitudes obey BCJ relations at each order in $\ap$ up to the order considered. Extending the method to higher orders is straightforward but introduces computational challenges as the number of terms grows considerably, as well as the multiplicity of the amplitudes which need to be computed to verify consistency with BCJ relations. We have found that in this representation of the Lagrangian, the operators proportional to $F^4$ can be written in a compact manner at all orders in $\ap$. We have reproduced some computations for this class of operators in a diagrammatic fashion, and we hope that further exploration of this method can shed some light on the overall structure of the effective Lagrangian.

Finally, we want to comment on the fact that the paper~ \cite{Johansson:2017srf} studied a further extension of the $(DF)^2 + \text{YM}$ theory which includes a massless biadjoint scalar $\phi^{aA}$ with a cubic interaction. This theory, named $(DF)^2 + \text{YM} + \phi^3$, reduces to $\text{YM} + \phi^3$ \cite{Chiodaroli:2014xia} in the $\ap \rightarrow 0$ limit. In ref.~\cite{Azevedo:2018dgo}, it was conjectured that the $(DF)^2 + \text{YM} + \phi^3$ theory is involved in the double copy expression of the heterotic string according to the schematic formula
\begin{equation}
M_{\text{het}}  = A_{(DF)^2 + \text{YM} + \phi^3} \underset{\mathrm{KLT}}{\otimes}
\, \text{sv} (A_{\text{open}})
\end{equation}
where on the right-hand side the single-valued map \cite{Schlotterer:2012ny, Stieberger:2013wea, Stieberger:2014hba} acts on the MZVs present in the open superstring amplitude expansion. We expect that the method described in this paper could be applied to the $(DF)^2 + \text{YM} + \phi^3$ Lagrangian to integrate out the massive states and obtain $\ap$ corrections to the $\text{YM} + \phi^3$ Lagrangian. We hope that progress on the kinematic algebra and a Lagrangian-level double copy \cite{Monteiro:2011pc, Cheung:2016prv, Chen:2019ywi, Borsten:2020xbt, Borsten:2020zgj, Chen:2021chy, Cheung:2021zvb, Brandhuber:2021bsf, Diaz-Jaramillo:2021wtl, Ben-Shahar:2021zww, Bonezzi:2022yuh, Chen:2022nei, Brandhuber:2022enp, Borsten:2022ouu, Bonezzi:2022bse, Ben-Shahar:2022ixa, Chen:2023ekh} can expand this program to provide insight on the effective actions for closed string theories.

\section*{Acknowledgments} 

It is a pleasure to thank Oliver Schlotterer for his encouragement during this project, as well as collaboration during its early stages. We also thank Henrik Johansson for helpful support and giving feedback on the draft, and Fei Teng for discussions. AG is supported by the
Black Hole Initiative and the Society of Fellows at Harvard University, as well as the Department
of Energy under grant DE-SC0007870. LG is supported in part by the Knut and Alice Wallenberg Foundation under grants KAW 2018.0116 ({\it From Scattering Amplitudes to Gravitational Waves}) and KAW 2018.0162 ({\it Exploring a Web of Gravitational Theories through Gauge-Theory Methods}).

\appendix 

\section{$F^4$ Lagrangian at all orders}
The non-local expression for the $F^4$ Lagrangian \eqref{eq:nonloc} can be used to write down an all orders expression in terms of the field strength and its derivatives. We need to consider separately the $\mathcal{O}(\ap^2)$ case since only the first term contributes:
\begin{equation}
    \mathcal{L}_{(2)}|_{F^4} = -\frac{1}{2} J^2,
\end{equation}
while for $n>2$ we can write
\begin{equation} \label{eq:f4_sources}
    \mathcal{L}_{(n)}|_{F^4} = -\frac{1}{2} J^\alpha D^{2(n-2)} J^\alpha + \frac{1}{2} K^\mu D^{2(n-3)} K_\mu.
\end{equation}
This can be further simplified by substituting the expressions for the sources. For simplicity we will write them as follows, using the numerical subscript notation for pairs of contracted field strengths:
\begin{align}
J^\alpha &= - \frac{C^{\alpha ab}}{2} F^a_1 F^b_1, \\
 K^{a \, \mu} &=  f^{abc} (D^\mu F_1^b) F^c_1.
\end{align}
The idea is to rewrite \eqref{eq:f4_sources} and manipulate the $J$ dependent term such that its kinetic piece is identical to that of the $K$ term. Starting from
\begin{align}
    \mathcal{L}_{(n)}|_{F^4}  &= -\frac{1}{8} C^{\alpha ab} C^{\alpha cd} F_1^a F_1^b (D_\mu D^\mu D^{2(n-3)} F_2^c F_2^d) \notag \\
    & \quad + \frac{1}{2} f^{abe} f^{ecd} (D_\mu F^a_1) F_1^b (D^{2(n-3)} (D^\mu F_2^c) F_2^d)
 \end{align}
where we have separated a pair of contracted derivatives on the first line, we can let one derivative act to its left, while the other commutes through all the $D^{2(n-3)}$ derivatives to act on the second pair of $F$'s. We can ignore the terms proportional to $F^5$ which arise from commuting the derivatives since we are only interested in $F^4$ operators. The two terms can then be combined by using the color factor relation \eqref{eq:col_4} and the Jacobi identity, giving a single term expression for the $F^4$ terms at any order in $\ap$, for $n>2$:
 \begin{equation} \label{eq:lag_f4_full}
 \mathcal{L}_{(n)}|_{F^4} = f^{ade}f^{ecb} (D_\mu F^a_1) F_1^b (D^{2(n-3)} (D^\mu F_2^c) F^d_2).
 \end{equation}
 This expression reproduces the $F^4$ terms we previously computed for $n=3$ in \eqref{eq:lag_ap3} and $n=4$ in \eqref{eq:lag_ap4_f4} and we can use it to spell out the $F^4$ terms at $\mathcal{O}(\ap^5)$ in trace form as follows:
\begin{align}
 \mathcal{L}_{(5)}|_{F^4} &= \text{Tr} \Big( 3 [D_\mu F_1, D_\rho D_\nu F_2][D^\mu F_2, D^\rho D^\nu F_1] + [D_\rho D_\nu D_\mu F_1, F_2] [D^\rho D^\nu D^\mu F_2, F_1]  \notag \\
 & \quad \quad + 2 ([D_\rho D_\mu F_1, D^\rho D_\nu F_2][D^\mu F_2, D^\nu F_1] + [D_\nu D_\rho F_1, D^\nu F_2][D_\mu D^\rho F_2, D^\mu F_1] \notag \\
 & \quad \quad \quad  + [D_\nu D_\mu F_1, D_\rho D^\nu F_2][D^\mu F_2, D^\rho F_1] + [D_\rho D_\nu D_\mu F_1, D^\rho D^\nu F_2][D^\mu F_2, F_1] \notag \\
 & \quad \quad \quad + [D_\rho D_\nu D_\mu F_1, D^\nu F_2][D^\rho D^\mu F_2, F_1] + [D_\rho D_\nu D_\mu F_1, D^\rho F_2][D^\nu D^\mu F_2, F_1]) \Big).
 \end{align}
 Of course there is some ambiguity in \eqref{eq:lag_f4_full} given by the possibility of integrating by parts. In the expression above, we have chosen to distribute the derivatives evenly between the two pairs of $F$'s.

\section{Expanded expressions for the Lagrangian and EOMs}

Starting from the expression for the Lagrangian \eqref{eq:lag2} we now expand all the tilded quantities to make the $B^\mu$ dependence of each term manifest. We also integrate by parts a piece of the $\tilde{F}^3$ term, which can be simplified in terms of a $K^\mu$ source dependent term.
\begin{align}
    \label{eq:lagfull}
    \mathcal{L} &= \frac{F^2}{4} -\frac{B^2}{2 \ap} + \frac{1}{2} B^a_\mu D^2 B^{a \,\mu} + \frac{\varphi^2}{2} - F^a_{\mu \nu} [B^\mu, B^\nu]^a + 2 (D_\mu B^a_\nu) [B^\mu, B^\nu]^a - \frac{3}{4} [B^\mu, B^\nu]^2 \notag \\
    & \quad + \ap \left( \frac{F^3}{3} - \frac{1}{2} \varphi^\alpha D^2 \varphi^\alpha - J^\alpha \varphi^\alpha + K^a_\mu B^{a \, \mu}  + 2 F^a_{\mu \nu} [D^\nu B^\rho, D_\rho B^\mu]^a +F^a_{\mu \nu} [D^\mu B^\rho, D^\nu B_\rho]^a  \right. \notag \\
    & \quad \quad \quad + F^a_{\mu \nu} [D^\rho B^\mu, D_\rho B^\nu]^a + 2 F^a_{\mu \nu} [D^2 B^\mu, B^\nu]^a +2 C^{\alpha ab} F^a_{\mu \nu} (D^\mu B^{b \, \nu}) \varphi^\alpha \notag \\
    & \quad \quad \quad + \frac{2}{3} (D_\mu B^a_\nu) [D^\nu B^\rho, D_\rho B^\mu]^a + 2 D_\mu B^a_\nu [D^\mu B^\rho, D^\nu B_\rho]^a - [F^{\mu \nu}, F_{\nu \rho}]^a [B^\rho, B_\mu]^a \notag \\ 
    & \quad \quad \quad + 4 [B^\mu, F_{\mu \rho}]^a [B_\nu, F^{\nu \rho}]^a + C^{\alpha ab} (D_\mu B^a_\nu) (D^\mu B^{b \, \nu}) \varphi^\alpha + C^{\alpha ab} (D_\mu B^a_\nu) (D^\nu B^{b \, \mu}) \varphi^\alpha \notag \\
    & \quad \quad \quad + (T^a)^{\alpha \beta} B^a_\mu \varphi^\alpha D^\mu \varphi^\beta + \frac{1}{6} d^{\alpha \beta \gamma} \varphi^\alpha \varphi^\beta \varphi^\gamma - 2 [F_{\mu \nu}, D^\rho B^\nu]^a [B_\rho, B^\mu]^a \notag \\
    & \quad \quad \quad + 4 [F_{\mu \nu}, B^\rho]^a [D_\rho B^\nu, B^\mu]^a - 2 [F_{\mu \nu}, B^\rho]^a [D^\nu B_\rho, B^\mu]^a + C^{\alpha ab} F^a_{\mu \nu} [B^\mu, B^\nu]^b \varphi^\alpha \notag \\
    & \quad \quad \quad - 2 [D_\mu B_\nu, D^\nu B^\rho]^a [B_\rho, B^\mu]^a + [D_\mu B_\nu, D^\rho B^\nu]^a [B_\rho, B^\mu]^a - [D_\mu B_\nu, D^\mu B^\rho]^a [B^\nu, B_\rho]^a \notag \\
    & \quad \quad \quad - 2 C^{\alpha ab} (D_\mu B^a_\nu) [B^\mu, B^\nu]^b \varphi^\alpha - \frac{1}{2} (T^a)^{\alpha \beta} (T^b)^{\beta \gamma} B^a_\mu B^{b \, \mu} \varphi^\alpha \varphi^\gamma \notag \\
    & \quad \quad \quad + [F_{\mu \nu}, [B^\nu, B^\rho]]^a [ B_\rho, B^\mu]^a -2 [F_{\mu \nu}, B^\mu]^a [[B^\rho, B^\nu], B_\rho]^a \notag \\
    & \quad \quad \quad + 2 [D_\mu B_\nu, [B^\nu, B^\rho]]^a [B_\rho, B^\mu]^a + \frac{1}{2} C^{\alpha ab} [B_\mu, B_\nu]^a [B^\mu, B^\nu]^b \varphi^\alpha \notag \\
    & \quad \quad \quad  \left. - \frac{1}{3} [B_\mu, B_\nu]^a [ [ B^\nu, B^\rho], [B_\rho, B^\mu]]^a \right).
\end{align}
For reference, we use the following identities for varying the fields:
\begin{subequations} \label{eq:varying}
\begin{align}
\frac{\delta F^{\mu \nu,a}}{\delta A^{\lambda,e}} \mathcal{O}^a &= -(D^\mu \mathcal{O}^e)\delta^\nu_\lambda +(D^\nu \mathcal{O}^e) \delta^\mu_\lambda,  \\
\frac{\delta \tilde{F}^{\mu \nu,a}}{\delta A^{\lambda,e}} \mathcal{O}^a &= \frac{\delta \tilde{F}^{\mu \nu,a}}{\delta B^{\lambda,e}} \mathcal{O}^a = - (\tilde{D}^\mu \mathcal{O}^e)\delta^\nu_\lambda + (\tilde{D}^\nu \mathcal{O}^e) \delta^\mu_\lambda,  \\
\frac{\delta (\tilde{D}^\mu \mathcal{O}_1^a)}{\delta A^{\lambda,e}} \mathcal{O}_2^a &= \frac{\delta (\tilde{D}^\mu \mathcal{O}_1^a)}{\delta B^{\lambda,e}} \mathcal{O}_2^a = -(\tilde{D}^\mu \mathcal{O}_2^a) \delta \mathcal{O}_1^a  -f^{eab} \mathcal{O}_1^a \mathcal{O}_2^b \delta^\mu_\lambda , \\
\frac{\delta (\tilde{D}^\mu \mathcal{O}_1^\alpha)}{\delta A^{\lambda,e}} \mathcal{O}_2^\alpha &= \frac{\delta (\tilde{D}^\mu \mathcal{O}_1^\alpha)}{\delta B^{\lambda,e}} \mathcal{O}_2^\alpha  = -(\tilde{D}^\mu \mathcal{O}_2^\alpha) \delta \mathcal{O}_1^\alpha +(T^e)^{\alpha \beta} \mathcal{O}_1^\alpha \mathcal{O}_2^\beta \delta^\mu_\lambda.
\end{align}
\end{subequations}
The expanded expressions for the equations of motion of the fields $\varphi^\alpha$ and $B^{a \, \lambda}$ read
\begin{align}
    \label{eq:phiexp}
     \varphi^\alpha &= \alpha \Big( D^2 \varphi^\alpha + J^\alpha -2 C^{\alpha ab} F^a_{\mu \nu} D^\mu B^{b \, \nu} - C^{\alpha ab} (D_\mu B^a_\nu) D^\mu B^{b \, \nu} + C^{\alpha ab} (C_\mu B^a_\nu) D^\nu B^{b \, \mu} \notag \\
     & \quad \quad - 2 (T^a)^{\alpha \beta} B^a_\mu D^\mu \varphi^\alpha - \frac{1}{2}d^{\alpha \beta \gamma} \varphi^\beta \varphi^\gamma + C^{\alpha ab} F^a_{\mu \nu} [B^\mu, B^\nu]^b + 2 C^{\alpha ab} (D_\mu B^a_\nu) [B^\mu, B^\nu]^b \notag \\
     & \quad \quad + (T^a)^{\alpha \beta} (T^b)^{\beta \gamma} B^a_\mu B^{b \, \mu} \varphi^\gamma - \frac{1}{2} C^{\alpha ab} [B_\mu, B_\nu]^a [B^\mu , B^\nu]^b \Big),
\end{align}
\begin{align}
    \label{eq:bexp}
   B^{a \, \lambda} &= \ap ( D^2 B^{a \, \lambda} - 2[B_\mu, D^\mu B^\lambda]^a + [B_\mu, [B^\mu, B^\lambda]]^a -2 C^{\alpha ab} B^{b \, \lambda} \varphi^\alpha ) \notag \\
   & \quad + \ap^2 \Big( K^{a \, \lambda} + 2 [D^\lambda D_\mu B_\nu, F^{\mu \nu}]^a + 2 [D^\lambda F_{\mu \nu}, D^\mu B^\nu]^a + 2 C^{\alpha ab} F^{b \, \lambda \mu} D_\mu \varphi^\alpha \notag \\
   & \quad \quad \quad \quad  +2 [D^\lambda D_\mu B_\nu, D^\mu B^\nu]^a - 2 [D^\lambda D_\mu B_\nu, D^\nu B^\mu]^a -2 [[B^\lambda, F_{\mu \nu}], F^{\mu \nu}] \notag \\
   & \quad \quad \quad \quad +2 C^{\alpha ab} (D^\lambda B^b_\mu) D^\mu \varphi^\alpha - 2 C^{\alpha ab} (D_\mu B^{b \, \lambda}) D^\mu \varphi^\alpha  - (T^a)^{\alpha \beta} (D^\lambda \varphi^\alpha)\varphi^\beta \notag \\
   & \quad \quad \quad \quad -2 [[D^\lambda B_\mu, B_\nu], F^{\mu \nu}]^a + [D^\lambda F_{\mu \nu}, [B^\mu, B^\nu]]^a -4 [[B^\lambda, F_{\mu \nu}], D^\mu B^\nu]^a \notag \\
   & \quad \quad \quad \quad-4 [[B^\lambda, D_\mu B_\nu], F^{\mu \nu}]^a -2 [D^\lambda D_\mu B_\nu, [B^\mu, B^\nu]]^a - 2 [[D^\lambda B_\mu, B_\nu], D^\mu B^\nu]^a \notag \\
   & \quad \quad \quad \quad+2 [[D^\lambda B_\mu, B_\nu], D^\nu B^\mu]^a - 4 [[ B^\lambda, D_\mu B_\nu], D^\mu B^\nu]^a + 4 [[B^\lambda, D_\mu B_\nu], D^\nu B^\mu]^a \notag \\
   & \quad \quad \quad \quad -2 C^{\alpha ab} [B^\lambda, B_\mu]^b D^\mu \varphi^\alpha - (T^a)^{\alpha \beta} (T^b)^{\beta \gamma} B^{b \, \lambda} \varphi^\alpha \varphi^\gamma + 2 [[B^\lambda, F_{\mu \nu}], [B^\mu, B^\nu]]^a \notag \\
   & \quad \quad \quad \quad +2 [[B^\lambda, [B_\mu, B_\nu]], F^{\mu \nu}]^a + 2 [[D^\lambda B_\mu, B_\nu], [B^\mu, B^\nu]]^a + 4 [[B^\lambda, D_\mu B_\nu], [B^\mu, B^\nu]]^a \notag \\
   & \quad \quad \quad \quad +4 [[B^\lambda, [B_\mu , B_\nu]], D^\mu B^\nu]^a -2 [[B^\lambda, [B_\mu, B_\nu]], [B^\mu, B^\nu]] \Big).
\end{align}

\bibliographystyle{JHEP}
\bibliography{bib}

\end{document}